\documentclass[12pt,article]{iopart}
\usepackage{amsmath}
\usepackage{amssymb}
\usepackage{rotate}
\usepackage{epsfig}
\usepackage{mathfrak}
\usepackage{bookmath}
\usepackage{feynmp}
\begin{document}
\title{\large Vortices in chiral, spin-triplet superconductors and superfluids}
\author{J.A. Sauls}                                                             
\address{Department of Physics \& Astronomy, Northwestern University \\ Evanston, IL 60208, USA}                                                                  \ead{\mailto{sauls@northwestern.edu}}  
\author{M. Eschrig}                                                             
\address{Institut f{\"u}r Theoretische Festk{\"o}rperphysik and DFG-Center for Functional Nanostructures,                                   Universit{\"a}t Karlsruhe, D-76128 Karlsruhe, Germany}                          
\ead{\mailto{eschrig@tfp.uni-karlsruhe.de}}                   
\date{January 15, 2009; revision: May 15, 2009}
\begin{abstract}
Superconductors exhibit {\sl unconventional} electronic and magnetic properties if the Cooper pair wave function breaks additional symmetries of the normal phase. Rotational symmetries in spin- and orbital spaces, as well as discrete symmetries such as space and time inversion, may be spontaneously broken. When this occurs in conjunction with broken global U(1) gauge symmetry, new physical phenomena are exhibited below the superconducting transition that are characteristic of the broken symmetries of the pair condensate.  This is particularly true of vortices and related defects. 
Superconductors with a multi-component order parameter exhibit a variety of different vortex structures and closely related defects that are not possible in condensates belonging to a one-dimensional representation. In this article we discuss the structure of vortices in Fermionic superfluids and superconductors which break chiral symmetry, i.e. combined broken time-inversion and 2D parity. In particular, we consider the structure of vortices and defects that might be realized in thin films of \Hea\ and the layered superconductor \sro, and identify some of the characteristic signatures of broken chiral symmetry that should be revealed by these defects.
\end{abstract} 
\pacs{67.30.he,67.30.hr,74.70.Pq,74.70.Tx}
\maketitle
\tableofcontents
\markboth{Vortices in chiral, spin-triplet superconductors and superfluids}{Vortices in chiral, spin-triplet superconductors and superfluids}
\section{Introduction}

The BCS theory of superconductivity \cite{bar57}, combined with refinements over several decades \cite{and58a,gor58,abr59b,eil68}, ranks among the major achievements in theoretical physics during the last century. The central feature of BCS theory is {\sl Cooper pair condensation} - i.e. the macroscopic occupation of a single quantum state of bound pairs of fermions. In its simplest form for s-wave, spin-singlet pairing the "order parameter", or Cooper pair amplitude,  is given a complex scalar function of the relative coordinate, $\vr$, and center of mass position, $\vR$, of the pairs,
\be\label{eq-order_parameter_s-wave}
\Psi  
= 
\langle \psi_{\uparrow}(\vR+\vr/2)\psi_{\downarrow}(\vR-\vr/2)\rangle 
= 
\vert\Psi\vert\,e^{i\vartheta}
\,.
\ee
The Cooper pair amplitude is both a measure of the condensate density, $|\Psi|^2 \sim{\cal O}(N/V)$, and a reflection of the \textsl{spontaneously broken symmetry} of the ordered phase, which in this case is the global U(1) gauge symmetry generated by Fermion particle number \cite{and66}. For homogeneous states the phase, $\vartheta$, is the signature of the broken U(1) symmetry, i.e. a macroscopic fraction of Fermions condense into a two-particle state with the \underline{same} quantum phase.
	The broken global U(1) symmetry also implies a family of \textsl{degenerate} ground states related to one another by the elements of the symmetry group. Long-wavelength spatial variations of the phase form a branch of low-lying collective excitations, the Anderson-Bogoliubov phase mode \cite{and58,bogoliubov58}, with an acoustic dispersion relation. This is a classic example of a Goldstone mode that accompanies a broken continuous symmetry. In superconductors, Cooper pairs are charged particles with charge 2$e$, and the requirement of {\it local}  U(1) gauge symmetry leads to a massive
Goldstone boson via the Higgs mechanism, that corresponds to the longitudinal oscillation of the electromagnetic vector potential, and is a high energy plasmon excitation\footnote{A direct consequence of the Higgs mechanism is the Meissner effect in superconductors.}. In neutral superfluids the phase mode is observable as fourth sound, a collective excitation of the condensate which has a velocity that is determined by the "phase stiffness", $\rho_s\propto |\Psi{\text{eq}}|^2$, or superfluid density, where $\Psi{\text{eq}}$ is the equilibrium condensate amplitude.
This rigidity stabilizes the condensate against phase fluctuations, and is responsible for the Josephson effect \cite{sim01}, persistent currents \cite{pek84}, and flux quantization in superconductors. Indeed the standard signature of BCS condensation is the quantization of flux in units of $\Phi_0=hc/2e$, or the quantization of circulation in units of $\kappa_0=h/2m$ in a neutral pair condensate of Fermions of mass $m$. 

Quantized flux and quantized circulation are consequences of a fundamental constraint on the phase of the pair amplitude,
\be
\oint_{\text{\sl C}} d\vell\cdot\grad\vartheta = N_{\text{\sl C}}\,\times\,2\pi
\,,
\ee
where $N_{\text{\sl C}}=0\pm 1,\pm 2,\ldots$ is the winding number of the phase around a closed contour {\small\sl C} within the condensate.
This quantization condition reflects the physical requirement that the order parameter be a single-valued field, which for a complex scalar pair amplitude is simply that the phase return to its value modulo $2\pi$.
For a multiply connected geometry this constraint and the phase stiffness implies energy barriers separating states of different winding number, and thus to quantized persistent currents. 
In a simply connected geometry states with $N_{\text{\sl C}}\ne 0$ force the condensate to effectively become multiply connected as there is necessarily one or more phase singularities interior to the contour {\small\sl C}. These singularities, points in two dimensions or lines in three dimensions, form a spectrum of topologically stable "defects" of the order parameter, i.e. {\sl phase vortices}, labelled by their \textsl{winding number}. These topologically stable defects are often energetically stable, or metastable because of energy barriers separating states with different winding numbers. The local structure of these defects, the \textsl{vortex cores}, are determined in part by their topology - their winding number in this simple case, the local spectrum of fermionic excitations and their interactions, as well as scattering of the Fermions by impurities or other sources of disorder \cite{rai96}.

BCS superconductors exhibit \textsl{unconventional} properties if the order parameter breaks additional symmetries of the normal phase. Spin- and orbital rotational invariance, as well as space and time inversion, may be spontaneously broken.  When this occurs in conjunction with broken U(1) gauge symmetry, new collective excitations of the pair condensate \cite{hir89,sau00a}, novel heat and transport properties \cite{gra96a,gra00a}, unconventional vortices \cite{sal87,tok90} as well as point defects \cite{bal06,gra00} characteristic of the complex symmetry breaking are possible \cite{mineev99}. 
In this article we discuss the structure of vortices in superfluid \He\ and superconductors which break spin and orbital rotation symmetries, parity and/or time-inversion symmetry. This class of pairing states is believed to describe the ground states of thin films of \Hea \cite{vor03}, the low-temperature superconducting phase of \upt\ \cite{sau94}, and the layered superconductor \sro \cite{mac03}.

\section{Theoretical Formalism}

The superconducting order parameter can be defined in terms of the condensate amplitude for Cooper pairs,
\be
\Psi_{\alpha\beta}=\langle \psi_{\alpha}(\vr_1)\psi_{\beta}(\vr_2)\rangle
\,,
\ee
where $\psi_{\alpha}(\vr)$ is the fermion field operator for spin projection $\alpha$.
This amplitude is directly related to the anomalous Matsubara propagator \cite{AGD2},
\ber 
\fl \qquad 
\F_{\alpha\beta}(\vr_1,\tau_1,\vr_2,\tau_2)
&=& 
-\langle \T_{\tau}\psi_{\alpha}(\vr_1,\tau_1)\psi_{\beta}(\vr_2,\tau_2)\rangle
\nonumber
\\
&=&
-\langle \T_{\tau}\psi_{\alpha}(\vR+\vr/2,\tau)\psi_{\beta}(\vR-\vr/2,0)\rangle
\equiv \F_{\alpha\beta}(\vR,\vr;\tau)
\,,
\eer
expressed here in terms of the center-of-mass, $\vR$, and relative (orbital) coordinate, $\vr$, of the pair, and the (imaginary) time difference, $\tau=\tau_1-\tau_2$. $\T_{\tau}$ is the (imaginary) time-ordering operation for Fermions. Fourier transforming with respect to the orbital coordinate and time gives the anomalous propagator for pairs with orbital momentum $\vp$, center of mass position $\vR$,
\be
\F_{\alpha\beta}(\vp,\vR;\epsilon_n)
=
\int_0^{\beta}d\tau\,e^{i\epsilon_n\tau/\hbar}\int d\vr\,e^{-i\vp\cdot\vr/\hbar}\,\F_{\alpha\beta}(\vR,\vr;\tau)
\,,
\ee
and Matsubara energy, $\epsilon_n=(2n+1)\pi/\beta$ for $n=0,\pm1, \ldots$, where $\beta =1/k_BT$. The local spectral function for the pairs is obtained by analytic continuation to the real axis, $i\epsilon_n\rightarrow\epsilon+i0^{+}$ \cite{AGD2}.

The orbital radius of a Cooper pair is of order $\xi_0 = \hbar v_f/2\pi k_B T_c$, which is typically large compared with atomic scales, e.g. the Fermi wavelength, $\xi_0\gg \hbar/p_f$. In this limit we can factor out the short wavelength variations of the pair amplitude and compute directly the spatial variations of the pair amplitude and order parameter on mesoscopic length scales defined by the coherence length, $\xi_0$, the mean-free path, $\ell$ and/or the London penetration depth, $\lambda$. This separation occurs because the high-energy, $\epsilon \sim E_f$, short-wavelength, $|\vp|\sim p_f$, properties of the Fermi liquid are unaffected by the long-wavelength, low-energy pairing correlations to leading order in $\hbar/p_f\xi_0, \hbar/p_f\ell, k_B T_c/E_f$ etc.
Thus, we define the amplitude for Cooper pairs near the Fermi surface by integrating over the low-energy band of states near the Fermi level \cite{sau94a},
\be\label{anomalous_qc-progator}
\f_{\alpha\beta}(\vp_f,\vR;\epsilon_n) \equiv 
\int_{-\Omega_c}^{+\Omega_c}\,d\xi_{\vp}\,\F_{\alpha\beta}(\vp,\vR;\epsilon_n)
\,,
\ee
where $\vp_f$ is the Fermi momentum, $\xi_{\vp}=v_f(|\vp|-p_f)$ and $\Omega_c \ll E_f$ is the quasiparticle bandwidth. A detailed discussion of this procedure is given in Refs.~\cite{ser83,sau94a,rai94,esc99fluk,esc01a}.
A synopsis of the quasiclassical transport theory for equilibrium states of inhomogeneous superconductors, as well as the  methods used to calculate the order parameter and electronic structure of superconducting states associated with impurities, interfaces and vortices is described below.

The propagator for Cooper pairs with relative momenta on the Fermi surface given in Eq.~\ref{anomalous_qc-progator} is coupled to the low-energy propagator for fermionic {\sl quasiparticles} of the superconductor. The coupled equations for quasiparticles and Cooper pairs are formulated in terms of a $4\times\,4$ matrix propagator in the combined particle-hole (Nambu) and spin space, which  incorporates the the spin correlations that are required by the pairing correlations,
\be
\whG(\vp,\vR;\epsilon_n) 
=
-\int_0^{\beta}d\tau\,e^{i\epsilon_n\tau}\int d^3r\,e^{-i\vp\cdot\vr}
\;
\langle \T_{\tau}\hat\psi(\vr_1,\tau)\hat{\bar{\psi}}(\vr_2,0)\rangle
\,,
\ee
where $\hat\psi = (\psi_{\uparrow}\,,\,\psi_{\downarrow}\,,\,\psi^{\dag}_{\uparrow}\,,\,\psi^{\dag}_{\downarrow})$ is the four-component (Nambu spinor) Fermion field operator, $\vr_1=\vR+\vr/2$,  $\vr_2=\vR-\vr/2$. Note also that for imaginary time evolution,  $\bar{\psi}(\vr,\tau)\equiv\psi^{\dag}(\vr,-\tau)$.
The quasiclassical matrix propagator is then defined by integrating the over the low-energy band of states near the Fermi surface, $|\xi_{\vp}|\le\Omega_c$, with $k_B T_c\ll\Omega_c\ll E_f$,
\be
\whg(\vp_f,\vR;\epsilon_n)=\frac{1}{a}
\int_{-\Omega_c}^{+\Omega_c}\,d\xi_{\vp}\,\whtauz\whG(\vp,\vR;\epsilon_n)
=
\begin{pmatrix}
\hat\g & \hat\f \\ \hat{\bar\f} & \hat{\bar\g}
\end{pmatrix}
\,,
\ee
where $\whtauz$ is the third Pauli matrix in Nambu space, and we have renormalized by dividing by the spectral weight, $a$, of the normal-state quasiparticle resonance, $0<a<1$.
Its structure in Nambu space is represented by the quasiparticle propagators, $\hat\g$ and $\hat{\bar\g}$, and the pair propagators, $\hat\f$ and $\hat{\bar\f}$. Each of these propagators is a $2\times 2$ matrix in spin space denoted by small hats. The pair propagator naturally separates into spin-singlet (anti-symmetric) and spin-triplet (symmetric) amplitudes,
\be
\hat\f_{\alpha\beta}(\vp_f,\vR;\epsilon_n) 
= 
\mbox{f}_0(\vp_f,\vR;\epsilon_n)\,\left(i\sigma_y\right)_{\alpha\beta}
+
\vec{\mbox{f}}(\vp_f,\vR;\epsilon_n)\cdot\left(i\vec{\sigma}\sigma_y\right)_{\alpha\beta}
\ee
while the natural description for the quasiparticle propagator is in terms of spin-scalar and vector components,
\be
\hat\g_{\alpha\beta}(\vp_f,\vR;\epsilon_n) 
= 
\mbox{g}(\vp_f,\vR;\epsilon_n)\,\delta_{\alpha\beta}
+
\vec{\mbox{g}}(\vp_f,\vR;\epsilon_n)\cdot\vec{\sigma}_{\alpha\beta}
\,,
\ee
where $\vec{\sigma}=\left(\sigma_x,\sigma_y,\sigma_z\right)$ are the Pauli spin matrices.
The description in terms for four matrix propagators is convenient, but redundant. Fundamental symmetry relations associated with Fermion anti-symmetry and conjugation (i.e particle $\leftrightarrow$ hole) connect the two diagonal and off-diagonal propagators,
\ber\label{g_symmetry}
\bar{\hat\g}(\vp_f,\vR;\epsilon_n) &=& \hat\g(-\vp_f,\vR;\epsilon_n)^{\star}
						     =   \hat\g(-\vp_f,\vR;-\epsilon_n)^{\mbox{\tiny tr}}
\\
\bar{\hat\f}(\vp_f,\vR;\epsilon_n)  &=& \hat\f(-\vp_f,\vR;\epsilon_n)^{\star}
						     =  -\hat\f(\vp_f,\vR;-\epsilon_n)^{\dag}
\,.
\label{f_symmetry}
\eer
where $\hb^{\text{tr}}$ ($\hb^{\dag}$) is the transpose (adjoint) of $\hb$.

\subsection{Transport Equation}

Extensions of Landau's theory of normal Fermi liquids to include BCS pairing correlations \cite {gor60}, electron-phonon interactions \cite{mig58}, and disorder \cite{abr59} culminated in the equations of Eilenberger \cite{eil68} and Larkin and Ovchinnikov \cite{lar69}, formulated in terms of the Nambu matrix propagator, $\whg(\vp_f,\vR;\epsilon_n)$, obeying transport-type differential equations along classical trajectories defined by the Fermi momentum,\footnote{Extensions of the transport theory of Eilenberger to non-equilibrium superconductivity were formulated by Eliashberg \cite{eli72} and by Larkin and Ovchinnikov \cite{lar76}. The nonequilibrium transport theory is discussed our companion paper on vortex dynamics in this volume \cite{esc09}.}
\be\label{qc_transport_equation}
i\vv_f\cdot\gradR\,\whg
+
\Big[i\epsilon_n\hat\tau_3 - \whDelta - \whvSig\,,\, \whg\Big] = 0
\,,
\ee
where $\left[ \ldots \right]$ is the commutator in Nambu space between the matrix propagator, $\whg$, the (imaginary) time-development operator, $\whtauz\partial_{\tau}\leftrightarrow i\epsilon_n\whtauz$, 
and various internal and external fields that couple to quasiparticles and pairs. 
Eilenberger's transport equation is obtained by an energy- and momentum-space renormalization of Dyson's equation and the full many-body equations for the self-energy functional. 
The normalization information contained in Dyson's equation is absent in Eilenberger's equation, but is recovered in the form of a contraint \cite{eil68,lar69} that must be satisfied by the physically allowed solutions of the transport equation. The normalization constraint on the Matsubara propagator is
\be\label{normalization} 
\whg(\vp_f,\vR;\epsilon_n)^2 =-\pi^2\widehat{1} 
\,.
\ee 
This constraint can be obtained as a boundary condition which supplements the transport equation, and reflects the fact that in the absence of any disturbance the propagator reduces to that for local homogeneous equilbrium. For more detailed discussions of the normalization condition see Refs. \cite{eil68,lar69,she85}.

The renormalization procedure leads to an expansion of the full many-body Green's functions and self-energies in terms of ratios of low to high energy scales, e.g. $k_BT_c/E_f$, $u_{\mbox{\tiny ext}}/E_f$, or short to long wavelength scales, e.g. $\hbar/p_f\xi_0$, $\hbar q/p_f$. We use a single parameter, $\sml$, to classify the order of magnitude of various terms in the transport equation, the magnitude of the self-energy and external fields, etc.
The results reported here are based on the leading order expansion of the self energy in $\sml$, particularly the mean-field pairing self-energy, $\whDelta$, that describes the inhomogeneous equilibrium state of the superconductor, and the effects disorder in terms of scattering by a dilute random distribution of impurities represented by the impurity self energy, $\whSig$. However, we omit the Landau molecular field self-energy, which is also of order $\sml$. We do not expect the Landau molecular field to lead us to qualitatively different conclusions regarding the basic structure of vortices in chiral p-wave states, however these terms may play a role in the relative stability of various multi-vortex configurations. We also treat the mean-field pairing self energy in the weak-coupling limit. 
Indeed a useful operational definition for the order parameter is the mean field pairing self energy, 
\be
\whDelta =
\begin{pmatrix} 0 & \hat\Delta \\ \hat{\bar\Delta} & 0 \end{pmatrix}
\ee
where $\hat\Delta$ is the $2\times 2$ spin matrix order parameter with components $\Delta_{\alpha\beta}(\vp_f,\vR;\epsilon_n)$, loosely referred to as the "gap function", is
is both a measure of the pairing correlation energy and the broken symmetry of the superconducting state. The gap function is determined by the BCS self-consistency equation relating $\hat\Delta(\vp_f,\vR;\epsilon_n)$ to the Cooper pair amplitude, $\hat\f(\vp_f,\vR;\epsilon_n)$, and the interaction, $\hat\lambda(\vp_f,\epsilon_n,\vp_f',\epsilon_n')$, that provides the ``pairing glue''. In the weak-coupling limit the frequency dependence of the pairing interaction, and the pairing self-energy, is constant within the bandwidth $|\epsilon_n| \le \Omega_c$. The order parameter then satisfies the weak-coupling gap equation,
%
\begin{fmffile}{fmf_gapeq}
\fmfcmd{style_def fermion expr p = cdraw p;
	shrink (0.7);
	cfill (marrow (p, .5))
	endshrink;
	enddef;}

\fmfcmd{style_def full_fermion expr p = 
        draw_dbl_plain p; 
	shrink (1.0);
        cfill (marrow (p, .5))
	endshrink;
	enddef;}

\fmfcmd{style_def double_fermion expr p = cdraw p;
	shrink (0.7);
        cfill (reverse tarrow (p, .5));
        cfill (harrow (p, .5))
	endshrink;
	enddef;}

\fmfcmd{style_def full_fermion_widearrows expr p = cdraw p;
	shrink (1.6);
        cfill (reverse tarrow (p, .50));
        cfill (harrow (p, .50))
	endshrink;
	enddef;}

\fmfcmd{style_def pair_fermion expr p = cdraw p;
	shrink (0.6);
        cfill (tarrow (p, .55));
        cfill (tarrow (reverse p, .55))
	endshrink;
	enddef;}

\fmfcmd{style_def conj_pair_fermion expr p = cdraw p;
	shrink (0.6);
        cfill (harrow (p, .50));
        cfill (harrow (reverse p, .50))
	endshrink;
	enddef;}

\fmfcmd{style_def fermion_low expr p = cdraw p;
	shrink (0.6);
	cfill (marrow (p, .45))
	endshrink;
	enddef;}

\fmfcmd{style_def fermion_high expr p = draw_dots p;
	shrink (0.6);
	cfill (marrow (p, .45))
	endshrink;
	enddef;}

\fmfcmd{style_def small_arrow expr p = cdraw p;
        shrink (0.8);
        cfill (marrow (p, .50))
        endshrink;
        enddef;}

\fmfcmd{style_def double_arrow expr p = cdraw p;
        shrink (0.8); 
	cfill (marrow (p, .45)); 
	cfill (marrow (p, .55)) 
	endshrink; 
	enddef;}

\fmfcmd{style_def wiggly_arrow expr p = cdraw (wiggly p);
        shrink (0.8);
        cfill (arrow p)
        endshrink;
        enddef;}

\fmfcmd{style_def impurity expr p = draw_dashes p;
	enddef;}

\fmfcmd{style_def impurity_arrow expr p = draw_dashes p;
	shrink (0.7);
	cfill (marrow (p, .45))
	endshrink;
	enddef;}

\fmfcmd{vardef bar (expr p, len, ang) = ((-len/2,0)--(len/2,0)) 
					rotated (ang + angle direction length(p)/2 of p) 
					shifted point length(p)/2 of p 
        enddef; 
	style_def cut_fermion expr p = cdraw p; 
	ccutdraw bar (p, 3mm,  60) 
	enddef;}

\fmfcmd{style_def photon_arrow expr p = draw_wiggly p;
        shrink (1.0);
        cfill (marrow (p, 0.475))
        endshrink;
        enddef;}

\ber\label{gapequation}
\Delta_{\alpha\beta}(\vp_f\vR) &\equiv &\qquad
\hspace{0.5cm}
\parbox{60mm}{
\begin{fmfgraph*}(60,40)
\fmfstraight
\fmfforce{(0.05w,0.5h)}{f}
\fmfforce{(0.95w,0.5h)}{i}
\fmfforce{(0.25w,0.5h)}{v2}
\fmfforce{(0.75w,0.5h)}{v1}
\fmfv{d.sh=circle,d.f=1,d.si=1mm}{v1,v2}
\fmf{boson,left=1,tension=0.3}{v2,v1}
\fmf{pair_fermion,right=0}{v1,v2}
\fmf{plain}{i,v1}
\fmf{plain}{v2,f}
\fmflabel{$-\vp_f,\beta$}{i}
\fmflabel{$\vp_f,\alpha$}{f}
\end{fmfgraph*}
}
\hspace*{-2.25cm}\quad 
\nonumber \\
&=&N_f\int d^2\vp_f'\,
\lambda_{\alpha\beta,\gamma\rho}(\vp_f,\vp_f')
\,T\!\sum_{\epsilon_n'}^{|\epsilon_n'|\le\Omega_c}\f_{\gamma\rho}(\vp_f';\epsilon_n')
\eer
\end{fmffile}
%
\begin{fmffile}{fmf_pairpropagator}
\noindent where $\f =$
\begin{fmfgraph*}(30,5)
\fmfstraight\fmfleft{v2}\fmfright{v1}
\fmf{pair_fermion,right=0}{v1,v2}
\end{fmfgraph*}
is the Cooper pair propagator and 
\begin{fmfgraph*}(30,3)
\fmfstraight\fmfleft{v2}\fmfright{v1}
\fmf{boson,right=0}{v1,v2}
\end{fmfgraph*}
is the bosonic propagator responsible for the pairing glue. $N_f$ is the single-spin normal-state density of states at the Fermi level and the integration is an average over the Fermi surface normalized to $\int d^2\vp_f \equiv \int dS_{\vp_f}\, n(\vp_f) = 1$, where $n(\vp_f)$ is the normalized angle-resolved density of states on the Fermi surface.
\end{fmffile}

The impurity self-energy is a functional of the quasiclassical propagator and defined in terms of the self-consistent impurity scattering t-matrix,
\be\label{self-energy_impurity}
\whvSig(\vp_f,\vR;\epsilon_n) = n_s\, \whmft(\vp_f,\vp_f,\epsilon_n;\vR)
\,,
\ee
where $n_s$ is the density of impurity density and $\whmft(\vp_f,\vp_f,\epsilon_n;\vR)$ is the forward scatting limit of the self-consistent t-matrix describing multiple-scattering of quasiparticles and pairs by an impurity in the superconductor,
%
\begin{fmffile}{fmf_tmatrix}
\ber
\fl \qquad
\whmft(\vp_f,\vp_f';\epsilon_n) &\equiv & 
\parbox{100mm}{
\hspace*{0.25cm}
\begin{fmfgraph*}(20,20)
\fmftop{tf,t1,ti}
\fmfstraight
\fmfbottom{f,v1,i}
\fmf{plain}{i,v1,f}
\fmfdot{v1}
\fmfv{d.sh=cross,d.f=1,d.si=2mm}{t1}
\fmf{dashes}{t1,v1}
\end{fmfgraph*}
\hspace*{0.25cm}$=$\hspace*{0.25cm}
\begin{fmfgraph*}(20,20)
\fmftop{tf,t1,ti}
\fmfstraight
\fmfbottom{f,v1,i}
\fmf{plain}{i,v1,f}
\fmfdot{v1}
\fmfv{d.sh=cross,d.f=1,d.si=2mm}{t1}
\fmf{dots}{t1,v1}
\end{fmfgraph*}
\hspace*{0.25cm}$+$\hspace*{0.25cm}
\begin{fmfgraph*}(40,20)
\fmftop{tf,t1,ti}
\fmfstraight
\fmfbottom{f,v2,v1,i}
\fmf{double,width=1}{v1,v2}
\fmf{plain}{i,v1}
\fmf{plain}{v2,f}
\fmfv{d.sh=circle,d.f=1,d.si=1.5mm}{v1,v2}
\fmfv{d.sh=cross,d.f=2,d.si=2mm}{t1}
\fmf{dots}{t1,v1}
\fmf{dots}{t1,v2}
\end{fmfgraph*}
\hspace*{0.25cm}$+$\hspace*{0.25cm}
\begin{fmfgraph*}(40,20)
\fmftop{tf,t1,ti}
\fmfstraight
\fmfbottom{f,v3,v2,v1,i}
\fmf{double,width=1}{v1,v2}
\fmf{double,width=1}{v2,v3}
\fmf{plain}{i,v1}
\fmf{plain}{v3,f}
\fmfdot{v1,v2,v3}
\fmfv{d.sh=cross,d.f=1,d.si=2mm}{t1}
\fmf{dots}{t1,v1}
\fmf{dots}{t1,v2}
\fmf{dots}{t1,v3}
\end{fmfgraph*}
\hspace*{0.25cm}$\dots\quad $
}
\nonumber \\
&&
\nonumber
\\
&=& \hspace*{0.25cm}
\parbox{100mm}{
\begin{fmfgraph*}(20,20)
\fmftop{tf,t1,ti}
\fmfstraight
\fmfbottom{f,v1,i}
\fmf{plain}{i,v1,f}
\fmfdot{v1}
\fmfv{d.sh=cross,d.f=1,d.si=2mm}{t1}
\fmf{dots}{t1,v1}
\end{fmfgraph*}
\hspace*{0.25cm}$+$\hspace*{0.25cm}
\begin{fmfgraph*}(50,20)
\fmftop{tf,t1,ti}
\fmfstraight
\fmfbottom{f,v2,v1,i}
\fmf{double,width=1}{v1,v2}
\fmf{plain}{i,v1}
\fmf{plain}{v2,f}
\fmfdot{v1,v2}
\fmfv{d.sh=cross,d.f=1,d.si=2mm}{t1}
\fmf{dots}{t1,v2}
\fmf{dashes}{t1,v1}
\end{fmfgraph*}
}
\nonumber
\\
&&
\nonumber
\\
&=& \whmfu(\vp_f,\vp_f';\epsilon_n) 
+ N_f\int d^2\vp_f''\,\whmfu(\vp_f,\vp_f'';\epsilon_n)\,\whg(\vp_f'';\epsilon_n)\whmft(\vp_f'',\vp_f';\epsilon_n)
\label{t_matrix}
\eer
\end{fmffile}

%
\begin{fmffile}{fmf_g}
\noindent where $\whg(\vp_f,\epsilon_n) =$
\begin{fmfgraph*}(20,5)
\fmfstraight
\fmfleft{f}\fmfright{i}
\fmf{double}{i,f}
\end{fmfgraph*}
is the full matrix propagator and $\whmfu(\vp_f,\vp_f')$, represented by the cross and dotted line, is the matrix element of the impurity potential between normal state quasiparticles with momenta $\vp_f$ and $\vp_f'$ on the Fermi surface.
\end{fmffile}
We model the disorder in terms of non-magnetic impurities that scatter quasiparticles isotropically, i.e. we retain only the s-wave contribution to the impurity potential, thus, $\whmfu = u_0\,\whtauz$. The corresponding scattering phase shift in the s-wave channel is defined by $\delta_0 = \tan^{-1}(\pi N_fu_0)$. The t-matrix equation, and thus the impurity self-energy, can be expressed as the solution of the matrix equation,
\be\label{t-matrix_S}
\whmft(\epsilon_n)= \whmfu+N_f\,\whmfu\,\langle\whg(\vp_f'';\epsilon_n)\rangle\whmft(\epsilon_n)
\,,
\ee
where $\langle\whg(\vp_f'';\epsilon_n)\rangle$ is the Fermi-surface average of the propagator. In the normal state, the low-energy quasiclassical propagator reduces to $\whg_{\text{N}} = -i\pi\,\sgn(\epsilon_n)\whtauz$, and thus the t-matrix descbribing the scattering of normal state quasiparticles and quasiholes near the Fermi surface becomes,
\be\label{t-matrix_N}
\whmft_{\text{N}}(\epsilon_n)= \frac{1}{\pi N_f}\,\sin\delta_0\,e^{-i\sgn(\epsilon_n)\delta_0\whtauz}
\,.
\ee
In this model the disorder is characterized by the strength of impurity potential, $u_0$, the mean density of scatterers, $n_s$ and the density of states for quasiparticles at the Fermi surface. 
The mean lifetime of a quasiparticle on the Fermi surface in the normal state, $\tau_{\text{N}}$, is obtained by analytic continuation of the self-energy to the real axis to obtain the retarded self-energy, $\whSig^{\text{R}}_{\text{N}}(\epsilon) = \whSig_{\text{N}}(\epsilon_n\rightarrow \epsilon + i0^+)$,
\be
-\Im \Sig^{\text{R}}_{\text{N}}(\epsilon) = \frac{\hbar}{2\tau_{\text{N}}}  = \frac{n_s}{\pi N_f}\,\sin^2\delta_0
\,.
\ee
The mean free path, $\ell_{\text{N}} = v_f\tau_{\text{N}}$, reduces to the classic result, $\ell_{\text{N}} = 1/n_s\sigma$, where $\sigma$ is the total cross section for scattering of a quasiparticle off an impurity, $\sigma = (4\pi\hbar^2/p_f^2)\,\sin\delta_0^2$. In terms of the dimensionless cross section $\bar\sigma \equiv \sin^2\delta_0$, the weak scattering limit (``Born limit'') corresponds to $\bar\sigma \rightarrow 0$, while the strong scattering (``unitarity limit'') corresponds to $\bar\sigma = 1$. 

For {\sl any} inhomogeneous superconducting state, the relative importance of impurity scattering is determined by two parameters, (i) a pair-breaking parameter defined by ratio of the scale of the pair correlation length and the mean free path, $x=\xi_0/\ell_{\text{N}}$ and (ii) the strength of the impurity scattering potential as measured by the dimensionless cross section, $\bar\sigma$. For unconventional pairing states, such as chiral p-wave superconductors, disorder destroys the superconducting state at a critical level of disorder given by $x_c \simeq 0.28$. The results reported below are based on self-consistent calculations of the order parameter, current distribution and local density of states for the clean limit and for modest levels of disorder with $\ell_{\text{N}} = 10\xi_0$ in the Born limit.

\subsection{Observables}

To complete the theoretical framework for our calculations of the vortex structure for Fermi-liquid superconductors and superfluids we include expressions for the equilibrium current, and local density of states in terms of the propagator. In particular, we define the normalized angle-resolved local density of states (LDOS) by dividing out the the angle-resolved normal-state DOS at the Fermi level, $N_f\,n(\vp_f)$. Thus, the normalized local spectral density due to particle-hole coherence of the pair condensate and the formation of sub-gap excitations associated with scattering and spatial inhomogeneities of the order parameter is obtained from the retarded propagators for particle and hole excitations by analytic continuation to the real axis,
\be
{\cal N}(\vp_f,\vR;\epsilon) = -\frac{1}{2\pi}
\Im\Tr{\widehat{\tau}_{3}\,\whg^{\mbox{\tiny R}}(\vp_f,\vR;\epsilon)}
\,,
\ee
where the trace is over the $4\times 4$ Nambu space. All equilibrium properties are determined by this spectrum. In particular, the equilibrium charge current associated with a topological defect in a superconductor is given by,
\ber\label{charge_current}
\vj(\vR)&=&N_f\int d^2\vp_f\,\int_{-\infty}^{+\infty}\,d\epsilon
\left[e\vv_f(\vp_f)\right]\,{\cal N}(\vp_f,\vR;\epsilon)\,\left(2f(\epsilon)-1\right)
\\
&=&N_f\int d^2\vp_f\,\left[e\vv_f(\vp_f)\right]\,
T\sum_{\epsilon_n}\,\onehalf\Tr{\widehat{\tau}_3\whg(\vp_f,\vR;\epsilon_n}
\,,
\label{charge_current_Matsubara}
\eer
where $f(\epsilon)=1/(e^{\beta\epsilon}+1)$ is the Fermi distribution. 
The orbital magnetization density is then given by $\curl{\vm}=-\frac{1}{c}\,\vj(\vR)$. For equilibrium states it is convenient to calculate the total current by transforming to the Matsubara representation, as shown in Eq. \ref{charge_current_Matsubara}. However, if we are interested in the contributions to the current from specific regions of the excitation spectrum then the formulation in terms of the LDOS is more useful. In particular, the angle-resolved {\sl spectral current density} is defined as the net current density carried by states at $\pm\vp_f$ and energy $\epsilon$ \cite{rai96},
\be
\vj(\vp_f,\vR;\epsilon)=N_f
\left[e\vv_f(\vp_f)\right]\,\left[{\cal N}(\vp_f,\vR;\epsilon) - {\cal N}(-\vp_f,\vR;\epsilon)\right]
\,.
\ee 

\section*{Ricatti Equations}

The order parameter and self-energy must be determined self-consistently with the solution of the transport equation for the propagator. An efficient method for solving the transport equation is based on a parameterization for the propagator that automatically satisfies the normalization constraint, and thus eliminates the possibility of spurious solutions. The parameterization is defined by \cite{nag93,esc99,esc00,vor03}
\be
\whg = -i\pi \widehat{\N}
\begin{pmatrix} 
\hat{1} + \hgamma\hbgamma & 2\hgamma  
\\ 
-2\hbgamma & - 1 - \hbgamma \hgamma 
\end{pmatrix}
\,,
\ee
where the prefactor is given by
\be
\widehat{\N} =
\begin{pmatrix} 
(\hat{1} - \hgamma\hbgamma)^{-1} & 0 
\\ 
0 & (\hat{1} - \hbgamma\hgamma)^{-1}
\end{pmatrix}
\,,
\ee
The coherence amplitudes $\hgamma$ and $\hbgamma$ are $2\times2$ matrices in spin space which
obey Ricatti-type equations,
\ber
i\vv_f\cdot\grad \hgamma + 2i\epsilon_n \hgamma - \hgamma\hat{\bar\Delta}\hgamma - 2\hat{\mbox{\small $\Sigma$}}\hgamma + \hat\Delta &=& 0 
\,, \\
i\vv_f\cdot\grad \hbgamma - 2i\epsilon_n \hbgamma - \hbgamma\hat\Delta\hbgamma + 2\hat{\bar{\mbox{\small $\Sigma$}}} \hbgamma+ \hat{\bar\Delta} &=& 0
\,,
\eer
and are simply related to the particle- and hole-like projections of the off-diagonal propagator,
\be
\hgamma=-(i\pi-\hat\g)^{-1}\hat\f
\,,\quad
\hbgamma=(i\pi+\hat{\bar\g})^{-1} \hat{\bar\f}\,,
\ee
where the projection operators in Nambu space for particle-like (+) and hole-like (-) excitations are given by \cite{she85},
\be
\widehat{\P}_{+} = {1\over 2}\left( 1 + {\whg\over -i\pi}\right)^{-1}
\quad,\quad
\widehat{\P}_{-} = {1\over 2}\left( 1 - {\whg\over -i\pi}\right)^{-1}
\,.
\ee
In particular, $\widehat{\P}_{\pm}^2=\widehat{\P}_{\pm}$ and $\widehat{\P}_{+}\widehat{\P}_{-}=\widehat{\P}_{-}\widehat{\P}_{+}=0$ follow directly from the normalization condition. The physical interpretation of these projectors, and the Ricatti amplitudes, is easily established by comparing the action of the projection operators on a general Nambu spinor with the particle-like and hole-like solutions of the Bogoliubov or Andreev equations.
The fundamental symmetry relation in Eqs. \ref{g_symmetry}-\ref{f_symmetry} for the particle and hole components of the quasiclassical propagator also imply symmetry relations relating the two Ricatti amplitudes, 
\be
\hbgamma(\hat{\vp}_f,\vR;\epsilon_n)
=
\hgamma(-\hat{\vp}_f,\vR; \epsilon_n)^{\star} 
\,.
\ee
The utility of the Ricatti equations is that they provide an efficient approach to solving the transport equation. The Ricatti equations are easily integrated numerically because they have numerically stable solutions. The extension of this formalism applicable to vortex dynamics is described in a companion article for this special collection, Ref. \cite{esc09}, as well as Refs. \cite{esc99,esc00,esc01}.

\subsection{Calculational Methods}

\begin{figure}[h]
\centerline{
\epsfxsize0.5\hsize\epsffile{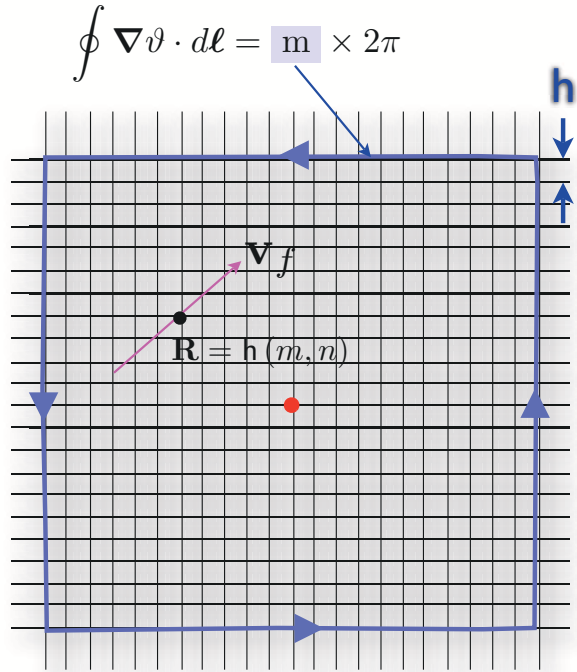}  
}
\caption{ 
\label{Fig_Grid}
Computational grid with spacing ${\mathsf h}$. At each grid point, labelled by $\vR={\mathsf h}(m,n)$, we integrate the transport equation along a classical trajectory (magenta) defined by the Fermi velocity, $\vv_f$. The global phase winding is enforced on the outer contour. The defect (red dot) is initially positioned at the center of the grid.
}
\end{figure}

The calculations for topological defects reported here are obtained from self-consistent solutions for the propagator, order parameter and self energy. For two-dimensional layered superconductors or superfluids we define all relevant functions, e.g. $\g(\vp_f, \vR, \epsilon_n)$, on a discrete set of points in both position and momentum space. The center of mass coordiate is discretized as $\vR = {\mathsf h}(m\va+m\vb)$, where ${\mathsf h}$ is the lattice spacing, $\va$ and $\vb$ are lattice unit vectors and $(m,n)$ is a pair of integers defining a discrete lattice point. A typical choice for the lattice spacing is ${\mathsf h}=0.25\,\xi_0$, and the dimensions of the largest grids used to carry out these calculations was $120\times 120$ lattice sites. The larger grids were used to investigate the stability of vortices multiple units of phase winding. We also used both rectangular and hexagonal lattices. For calculations of isolated defects we place the defect initially at the origin and enforce the constraint on the global phase winding by fixing the order parameter by its asymptotic form for $|\vR|\rightarrow\infty$ along an exterior contour as shown in Fig. \ref{Fig_Grid}.  At each grid point we use a fourth order Runge-Kutta algorithm to integrate the transport equation along a discrete set of trajectories representative of the Fermi surface and the basis functions defining the pairing symmetry. For the calculations reported here we used sixteen trajectories equally spaced in momentum space around the Fermi surface, which for simplicity we assumed to be cylindrical. However it is straight forward to implement more detailed Fermi surface geometries if needed (see Ref. \cite{fog04}). We also calculate the propagator for a finite set of Matsubara frequencies. The maximum Matsubara frequency required to achieve high precision increases with decreasing temperature. For $T/T_c  =0.4$ eight Matsubara frequencies is typically sufficient. 

A calculation is initialized by a ``seed'' for the order parameter field, with the topological constraint implemented on the boundary as described above. The propagator is then calculated for all the grid points, and for the set of Fermi surface trajectories and Matusbara frequencies at each grid point. The order parameter and impurity self energy are then calculated from discretized forms of Eqs. \ref{gapequation}, \ref{self-energy_impurity} and \ref{t-matrix_S}. The updated values of the order parameter and impurity self-energy are then used as new inputs to the transport equation, which is solved again for each grid point to obtain an improved solution for the propagator. This iterative procedure is continued until it converges to a specified precision for the deviations of the order parameter and self energy between iterations. The final results for the equilibrium order parameter and excitation spectrum are found to be insensitive to the initial order parameter field, except of course for the constraint on the phase winding. 

\subsection{Pairing Symmetry}

Fermi statistics requires that pair amplitude, $\f_{\alpha\beta}$, the pairing interaction, $\lambda_{\alpha\beta;\gamma\rho}$, and thus the gap function, $\Delta_{\alpha\beta}$, obey the anti-symmetry condition,\footnote{Unconventional pairing states which have no equal-time average, {\ie} `odd-frequency' pairing states, obey a more general anti-symmetry condition which includes the imaginary time coordinate \cite{esc07}. Such states have the parity assignments for total spin quantum numbers interchanged. See Berezinskii \cite{ber74}, and for a more recent discussion of these states, Balatsky and Abrahams \cite{bal92}.}
\be
\Delta_{\alpha\beta}(\vp_f,\vR)=-\Delta_{\beta\alpha}(-\vp_f,\vR)
\,.
\ee

For materials in which the normal metallic state possesses inversion symmetry the pairing interaction separates into even- (g) and odd-parity (u) channels, which are respectively anti-symmetric and symmetric under spin exchange and correspond to the spin-singlet ($S=0$) and spin-triplet ($S=1$) pairing channels,
\be
\lambda_{\alpha\beta;\gamma\rho}(\vp_f,\vp_f')
=
(i\sigma_y)_{\alpha\beta}\,\lambda^{(g)}(\vp_f,\vp_f')\,(i\sigma_y)_{\gamma\rho}
+
(i\sigma_y\vec{\sigma})_{\alpha\beta}
\cdot \overset{{\small \leftrightarrow}}{\lambda}^{(u)}(\vp_f,\vp_f') \cdot 
(i\vec{\sigma}\sigma_y)_{\gamma\rho}
\,,
\ee
Furthermore, the pairing interaction separates into a sum over invariant bilinear products of basis functions for the irreducible representations, $\Gamma$, of the point group, for both even- (${\eta}_{\Gamma\nu}(\vp_f)$) and odd-parity ($\vec{\eta}_{\Gamma\nu}(\vp_f)$) sectors, 
\ber
\lambda^{(g)}(\vp_f,\vp_f')
&=&
\sum_{\text{$\Gamma_g , \nu$}}
                           \lambda_{\text{$\Gamma$}}
                           \eta_{\text{$\Gamma\nu$}}(\vp_f)
                           \eta_{\text{$\Gamma\nu$}}^*(\vp_f')
\,,
\\
\overset{{\small \leftrightarrow}}{\lambda}^{(u)}(\vp_f,\vp_f')
&=&
\sum_{\text{$\Gamma_u , \nu$}}
                           \lambda_{\text{$\Gamma$}}
                           \vec{\eta}_{\text{$\Gamma\nu$}}(\vp_f)
                           \otimes
                           \vec{\eta}_{\text{$\Gamma\nu$}}^{\dag}(\vp_f')
\,.
\eer
These basis functions are the \textsl{eigenfunctions} of the linearized gap equation, while $\lambda_{\text{$\Gamma$}}$ is the eigenvalue that determines the instability temperature for Cooper pairs belonging to the the irreducible representation, $\Gamma$, i.e. $1/\lambda_{\Gamma} = \ln(1.13\Omega_c/ T^{\text{$\Gamma$}}_c)$. Except in rare cases of accidental degeneracy, or a weakly broken symmetry \cite{hes89}, the pairing interactions are well separated and the most attractive channel determines both the transition temperature, $T_c$ and the basis functions, $\{\eta_{\text{$\Gamma\nu$}}\}$, for the irreducible representation, $\Gamma$, that determines the pairing state(s) of the superconductor. The index $\nu$ labels the basis functions for the representation $\Gamma$ with dimension $d_{\text{$\Gamma$}}$. Although the exact basis functions depend on details of the anisotropic Fermi surface and pairing interaction, representative basis functions which exhibit all the broken symmetry properties of the irreducible representation are easily constructed \cite{yip93c}. Below we list the irreducible representations and \textsl{representative} basis functions for the group $D_{4h}$, appropriate for \sro in Table \ref{tab_basis}. 

%
\begin{table}[h] 
\label{tab_basis}
\centerline{
\begin{tabular}{|c|c|c|c|}
\hline
$\Gamma_g$ & $\eta_{\text{$\Gamma\nu$}}$ & $\Gamma_u$ & $\vec{\eta}_{\text{$\Gamma\nu$}}$ \\
\hline
A$_{1g}$ & 1 &
A$_{1u}$ & $\vec{d}\,\hat{\vp}_z$  \\
\hline
A$_{2g}$ & $\hat{\vp}_x \hat{\vp}_y(\hat{\vp}_x^2-\hat{\vp}_y^2)$ &
A$_{2u}$ & $\vec{d}\,\hat{\vp}_z \hat{\vp}_x \hat{\vp}_y(\hat{\vp}_x^2-\hat{\vp}_y^2)$  \\
\hline
B$_{1g}$ & $\hat{\vp}_x \hat{\vp}_y$ &
B$_{1u}$ & $\vec{d}\,\hat{\vp}_z \hat{\vp}_x \hat{\vp}_y$ \\
\hline
B$_{2g}$ & $(\hat{\vp}_x^2-\hat{\vp}_y^2)$ &
B$_{2u}$ & $\vec{d}\,\hat{\vp}_z\,(\hat{\vp}_x^2-\hat{\vp}_y^2)$  \\
\hline
E$_{g}$ & $\hat{\vp}_z \left(\begin{array}{c}\hat{\vp}_x + i \hat{\vp}_y \\ \hat{\vp}_x - i \hat{\vp}_y \end{array}\right)$ &
E$_{u}$ & $\vec{d}\left(\begin{array}{c}\hat{\vp}_x + i \hat{\vp}_y \\ \hat{\vp}_x - i \hat{\vp}_y\end{array}\right)$ \\
\hline
\end{tabular}
}
\caption{Representative basis functions for $D_{4h}$ appropriate to strong spin-orbit coupling with $\vec{d} || \hat\vc$, or negligible spin-orbit coupling and arbitrary direction of $\vec{d}$.}
\end{table}
\bigskip

Thus, barring accidental near degeneracy \cite{escsro}, the order parameter, $\Delta_{\alpha\beta}(\vp)$,  is defined by a single representation, and is either even- or odd-parity, and therefore spin-singlet or spin-triplet, respectively,\footnote{For materials in which spin-orbit effects are strong, i.e. actinide and rare earth heavy fermion metals,  the labels "spin-singlet" and "spin-triplet" do not refer simply to the eigenvalues of the spin operator for electrons, but rather a ``pseudo-spin''. The Kramers' degeneracy in zero-field guarantees that each $\vp$ state is two-fold degenerate. These two states may be labeled by a pseudo-spin quantum number $\alpha$. Thus, for many of our considerations the distinction between spin and pseudo-spin is not important \cite{and84a,vol85,lee86}.}
\ber
\Delta_{\alpha\beta}(\vp_f)
&=&
\Delta(\vp_f)\,(i\sigma_y)_{\alpha\beta}
\, ,\;\;\qquad {\rm singlet}\, (S=0)
\,,
\\
\Delta_{\alpha\beta}(\vp_f)
&=&
\vec{\Delta}(\vp_f)\cdot(i\vec{\sigma}\sigma_y)_{\alpha\beta}
\,,\qquad {\rm triplet}\, (S=1)
\,,
\eer
with $\Delta(\vp_f)=\Delta(-\vp_f)$ and $\vec{\Delta}(\vp_f)=-\vec{\Delta}(-\vp_f)$. The expansion in the eigenfunctions belonging to the dominant representation $\Gamma$ (either even or odd-parity) of the linearized gap equation gives,
\ber
\Delta(\vp_f)
&=&
\sum_{\nu}^{d_{\text{$\Gamma$}}}
\Delta_{\text{$\nu$}}
\,\eta_{\text{$\Gamma\nu$}}(\vp_f)
\,,
\\
\vec{\Delta}(\vp_f)
&=&
\sum_{\nu}^{d_{\text{$\Gamma$}}}
\Delta_{\text{$\nu$}}
\,\vec{\eta}_{\text{$\Gamma\nu$}}(\vp_f)
\,.
\eer

\subsection{Odd-Parity, Spin-Triplet Pairing}

The prototype for odd-parity, spin-triplet pairing is realized in the neutral Fermi superfluid \He. In particular the A-phase of superfluid \He\ is identified as an equal-spin pairing state with a p-wave order parameter of the form \cite{vollhardt90},
\be
\vec{\Delta}(\vp_f) = \frac{\Delta_{0}}{\sqrt{2}}\,\vec{d}\,\left(\vm + i \vn\right)\cdot\hat{\vp}
\,,
\ee
where $\vell = \vm\times\vn$ is the quantization axis for the orbital angular momentum of the pairs, i.e. $\vell\cdot\widehat{\vL}^{\text{orb}}\left(\vm + i \vn\right)\cdot\hat{\vp} = +\hbar \left(\vm + i \vn\right)\cdot\hat{\vp} $, $\hat{\vp}=\vp_f/|\vp_f| $and $\vec{d}$ is the axis along which the pairs have zero spin projection, i.e. $\vec{d}\cdot\widehat{\vS}_{\text{pair}}\vec{\Delta} = 0$. This corresponds to a spin-state that is an equal amplitude superposition of $\ket{\uparrow\uparrow}$ and $\ket{\downarrow\downarrow}$ spin states in the plane perpendicular to $\vec{d}$. In bulk \Hea, which is stable over a narrow temperature range at high pressures, both the orbital and spin quantization axes are broken symmetry directions, aligned only by external walls or weak symmetry breaking fields. In particular, the very weak nuclear dipolar energy is minimized by orienting the nuclear spins in the orbital plane, i.e. $\vec{d} || \pm \vell$ \cite{leg75}. In thin films of superfluid \He\ scattering by the surface and substrate leads to substantial pair-breaking \textsl{unless} the pairs are confined to the $x$-$y$ plane of the film \cite{amb75}. This condition favors the A-phase with $\vec{d}||\vell || \hat{\vz}$ over the B-phase for the entire pressure range (c.f. Fig. 3 in Ref. \cite{vor03}). However, in this reduced geometry the planar phase, which is also an equal-spin pairing state, but with $d_z = 0$ is \textsl{degenerate} with the A-phase in weak-coupling theory. It is unknown whether or not the A-phase or the planar phase is the stable phase in thin films, particularly at low pressures where strong-coupling effects are extremely small \cite{vor07}. For the planar state the two spin states are aligned out of the plane of film and correlated with the orbital pairing state to form a state with zero total angular momentum projection along $\hat{\vz}$,
\be
\vec{\Delta}(\vp_f) = \frac{\Delta_{0}}{2}\,
\left\{
\left(\hat{x} + i \hat{y}\right)\,\left(\hat{\vp}_x - i \hat{\vp}_y\right) 
+ 
\left(\hat{x} - i \hat{y}\right)\,\left(\hat{\vp}_x + i \hat{\vp}_y\right) 
\right\}
\,.
\ee

\subsection{Broken Time-Inversion and Chirality}

In contrast to the planar phase the A-phase spontaneously breaks both time-reversal symmetry and reflection symmetry in a plane normal to the film. This combination of broken 2D reflection symmetry and time-reversal symmetry is referred to as \textsl{broken chiral symmetry}. The structure of vortices in chiral superfluid \Hea\ are fundamentally different than vortices in non-chiral condensates such as the planar phase of superfluid \He.

There are also candidates for triplet pairing in the class of strongly correlated electronic superconductors. For example, UPt$_3$ is believed to have an odd-parity order parameter belonging to the E$_{2u}$ representation of D$_{6h}$ with a low-temperature phase that is an equal-spin pairing state with $\vec{d}||\hat{\vz}$, and breaks parity and time-reversal symmetry, i.e. $\vec{\Delta} = \vec{d}\,\hat\vp_z\,\left(\hat{\vp}_x + i \hat{\vp}_y\right)^2$.\footnote{The higher symmetry of the hexagonal rotation group  allows for a second two-dimensional E representation formed from the basis functions of the two B representations of D$_{4h}$. The odd-parity version,  E$_{2u}$, with $\vec{d}||\hat{\vz}$ has basis functions $\vec{d}\,\hat\vp_z\,(\hat{\vp}_x\pm i \hat{\vp}_y)^2$.} The evidence and analysis favoring this identification comes from observations of a multiple superconducting phases, low-temperature transport properties and observation of anisotropic Pauli limiting, and is summarized in Refs. \cite{cho91,sau94,gra00a}. More recent small-angle neutron scattering studies of the flux lattice transitions also support this identification \cite{hux00,cha01}.

The layered superconductor, \sro\ is also candidate for chiral spin-triplet pairing with a proposed order parameter that is essentially identical to that for a thin film of \Hea\ \cite{sig99}. Strong evidence for unconventional pairing in \sro\ is provided by the suppression of $T_c$ below 1.5 K by non-magnetic impurities \cite{mac98a}, while the absence of a Hebel-Slichter peak in $1/T_1T$ probed by NQR and Knight shift measurements \cite{ish98} are cited in support a spin-triplet order parameter with the spin-quantization axis $\vec{d} || \hat{\vz}$. Phase-sensitive measurements based on the current-phase relation for Josephson junctions coupling a conventional s-wave superconductor, Au-In alloy, and \sro\ support the interpretation that \sro\ is an odd-parity superconductor \cite{nel04}. Studies of the excess current obtained from point contact spectroscopy agree quantitatively with a $p$-wave triplet pairing state \cite{laube03}. Evidence for broken time-reversal symmetry of the superconducting order parameter is provided by $\mu$SR \cite{luk98} and magneto-optical Kerr effect \cite{xia07} experiments indicating that spontaneous currents and magnetic fields develop below $T_c$ in the Meissner state. However, the identification of broken chirality is controversial because local magnetic probes have not observed the fields associated with the domain-wall currents that are expected from broken chiral symmetry \cite{kir07}.

\section{Vortex structure in chiral p-wave, spin-triplet condensates}

Superconductors with a multi-component order parameter exhibit a variety of different vortex structures not possible in condensates belonging to a one-dimensional representation.  These structures can lead to multiple superconducting phases as well as novel electronic and magnetic properties \cite{tok90}.
Theoretical studies of these vortex structures can provide important information for future experimental studies of vortex phases in triplet superconductors.

Theoretical analysis of the stability of the topologically stable vortex structures requires calculations of the free energy of these structures, typically taking into account the interactions with other vortices in the vortex lattice structure.  Here we restrict our analysis and discussion to the internal structure of individual vortices appropriate to the low-field regime $H\simeq H_{c_1}$. 

For any inhomogeneous superconducting state, and specifically for a vortex in a p-wave, spin-triplet superconductor or superfluid, the order parameter can be expanded in basis functions of the relevant irreducible representation, $\Gamma$, of the symmetry group,
\be
\vec{\Delta}(\vp_f,\vR) = \sum_{\nu}\Delta_{\nu}(\vR)\vec{\eta}_{\text{$\Gamma,\nu$}}(\vp_f)
\,.
\ee
For chiral p-wave states appropriate for thin films of \Hea\ and \sro\ with $\vec{d}||\vec{\vz}$, the representation is two-dimensional and the basis functions, $\vec{\eta}_{\text{$\Gamma,\nu$}}(\vp_f) = \vec{d}\,\eta_{\text{$\nu$}}(\vp_f)$ with $\eta_{\pm}(\vp_f) = \hat{\vp}_x \pm i \hat{\vp}_y$, correspond to orbital angular momentum eigenstates with $L_z^{\text{orb}}=\pm\hbar$. For \Hea\ these basis functions are exact to the extent that higher-order representations, e.g. f-waves with $L_z=\pm 1$, can be negelected \cite{sau86}. In the case of \sro\ higher-order harmonics belonging to the same irreducible representation, $E_u$, are possible. Typically, we can neglect these higher-order terms as the basic features of the vortex structure are obtained with the lowest order odd-parity harmonics.

Because time-reversal symmetry is broken for chiral, p-wave superconductors, there are two possible degenerate ground states defined by their spontaneous orbital motion. The homogeneous ground state is spontaneously chosen between the two degenerate time-reversed states with $L_z^{\text{orb}}=\pm\hbar$ \cite{hes89}. In the following sections we consider vortices in a single domain with internal {\sl orbital} angular momentum $L_z^{\text{orb}}= + \hbar$. Vortices in the time-reversed ground state can be obtained simply by applying the time-inversion operation: $\vDelta \rightarrow \op{\mathsf T} * \vDelta$.

\subsection{Global and local winding numbers for vortices}

Consider an isolated vortex with circulation $m\times 2\pi$ far from the vortex core, with $m$ an integer.  In this limit the amplitude approaches its bulk value but carries the global phase winding,
\be\label{global}
\lim_{|\vR|\rightarrow\infty}\vec{\Delta}(\vp_f,\vR) 
\equiv 
\vec{\Delta}_{\infty}(\vp_f,\vR) 
= 
\vec{d}\,
\frac{\Delta_{0}}{\sqrt{2}}\,e^{i m\phi}\,\left(\hat{\vp}_x+i\hat{\vp}_y\right)
\,.
\ee
Asymptotically the vortex is cylindrically symmetric, and is thus an eigenfunction of the generator, $\widehat{L}_z$, for rotations about the $\hat\vz$ axis,\footnote{Spin rotations play no role for \Hea\ films with fixed $\vec{d}||\hat{\vz}$ .}
\be
\widehat{L}_z * \vec{\Delta}_{\infty}(\vp_f,\vR) = l\,\hbar\,\vec{\Delta}_{\infty}(\vp_f,\vR)
\,,\qquad l = 0,\pm 1,\pm 2, \ldots
\ee
where 
$\widehat{L}_z = \widehat{L}_z^{\text{cm}} + \widehat{L}_z^{\text{orb}}$ is the sum of the z-component of the angular momentum operator for the center of mass (CM) of the pairs, $\widehat{L}_z^{\text{cm}} = \frac{\hbar}{i}\partial/\partial\phi$ with $\vR=|\vR|(\cos\phi,\sin\phi)$, and the orbital angular momentum of the pairs relative to the CM, $\widehat{L}_z^{\text{cm}} = \frac{\hbar}{i}\partial/\partial\varphi_{\hat\vp}$, where $\varphi_{\hat\vp}$ is the azimuthal angle of the relative momentum of the pairs, $\vp_f = p_f (\cos\varphi_{\hat\vp}, \cos\varphi_{\hat\vp})$.
Thus, we have $l = m + 1$.  At finite, but large distances from the vortex core, i.e. $|\vR| \gg \xi_0$, gradients of the asymptotic vortex order parameter \textsl{induce} the degenerate, time-reversed pairing state leading to an order parameter of the more general form,
\be\label{local}
\vec{\Delta}(\vp_f,\vR) 
= 
\vec{d}\,
\Bigg[
|\Delta_{+}(\vR)|\,e^{i m\phi}\,\frac{\left(\hat{\vp}_x + i\hat{\vp}_y\right)}{\sqrt{2}}
+
|\Delta_{-}(\vR)|\,e^{i p\phi}\,\frac{\left(\hat{\vp}_x - i\hat{\vp}_y\right)}{\sqrt{2}}
\Bigg]
\,.
\ee
It is clear that the time-reversed phase with winding number $p$ is nucleated in the vicinity of the vortex core.  At large distances the vortex retains axial symmetry. This enforces a condition on the \textsl{local winding number,} $p$, of the time-reversed phase in the core, i.e. $p - 1 = m +1$ \cite{sau94}. This condition is exact for vortices in a pure p-wave chiral ground state such as \Hea. For chiral ground states in superconductors with discrete symmetry the condition on the local phase winding can be generalized. For example, for the discrete point group D$_{4h}$ appropriate to \sro\ the condition on the local phase winding becomes,
\be
p = m +2 + 4n
\,,
\ee
where $n=0, \pm 1, \pm 2, \ldots$ includes the effect of higher order harmonics comprising the ground state that are invariant under 4-fold rotations. If deviations from cylindrical symmetry are small we can neglect corrections coming from $n\ne 0$. For simplicity we restrict our quantitative analysis to vortices with $n=0$, but comment on the qualitative effects of the higher order harmonics where appropriate.

%
\begin{table}[h] 
\label{table-winding_numbers}
\centerline{
\begin{tabular}{ccc}
%
%
\begin{tabular}{|c|c|}
\hline
$\Delta_{+}$ & $\Delta_{-}$\\
\hline
\epsfxsize0.10\hsize\epsffile{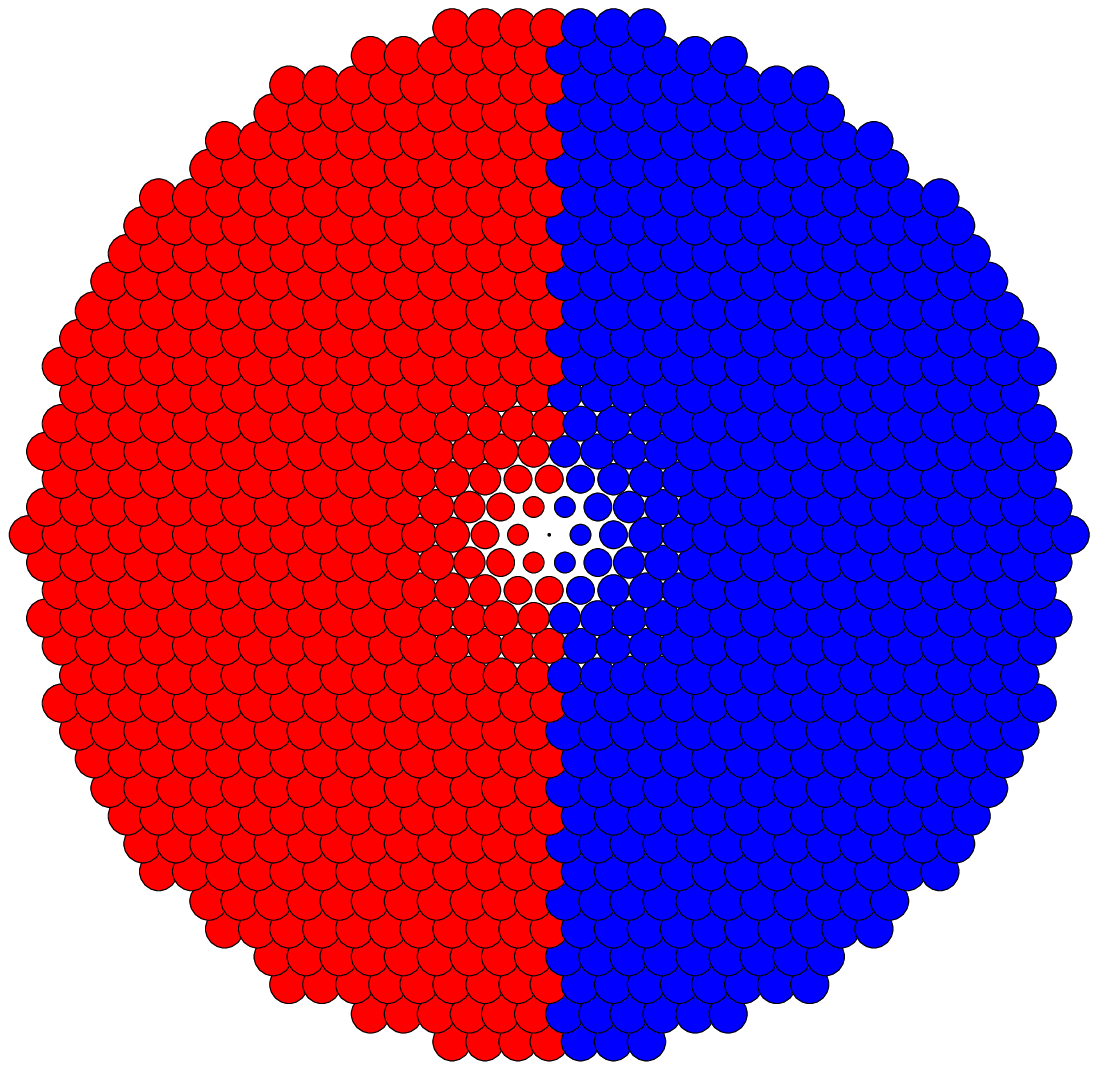} & \epsfxsize0.10\hsize\epsffile{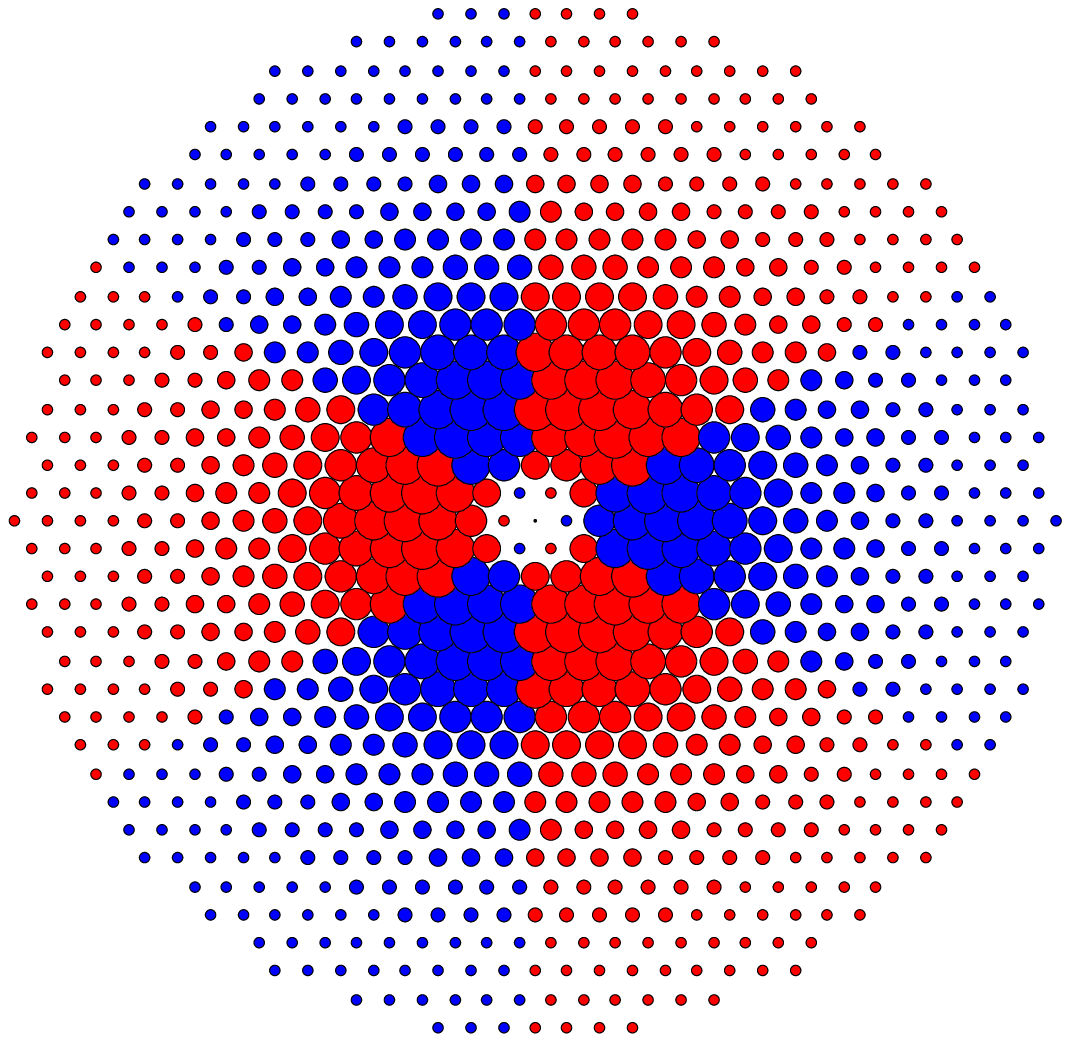} 
\\
$m= + 1$                                                & $p= + 3$
\\
\hline\hline
\epsfxsize0.10\hsize\epsffile{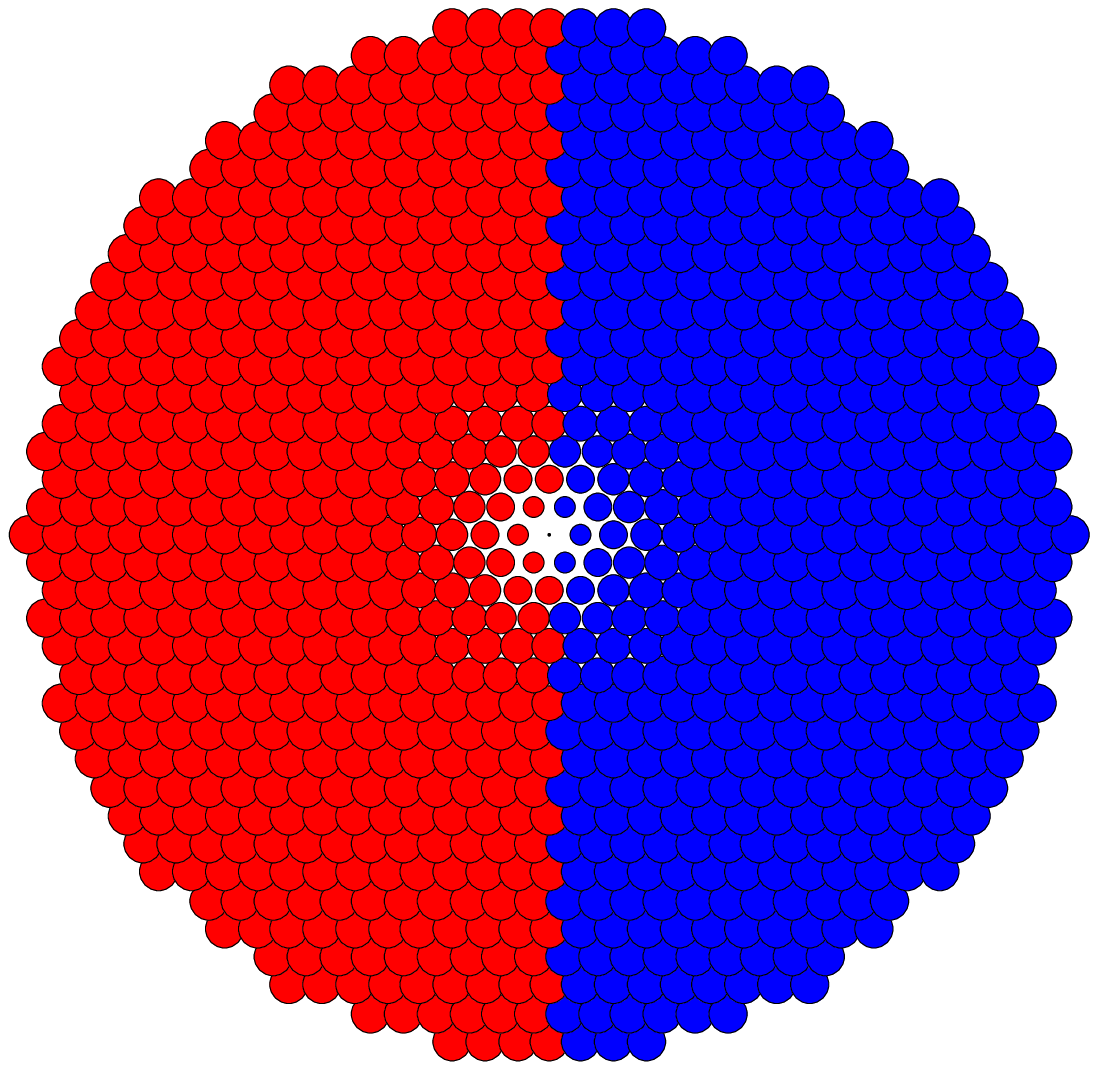} & \epsfxsize0.10\hsize\epsffile{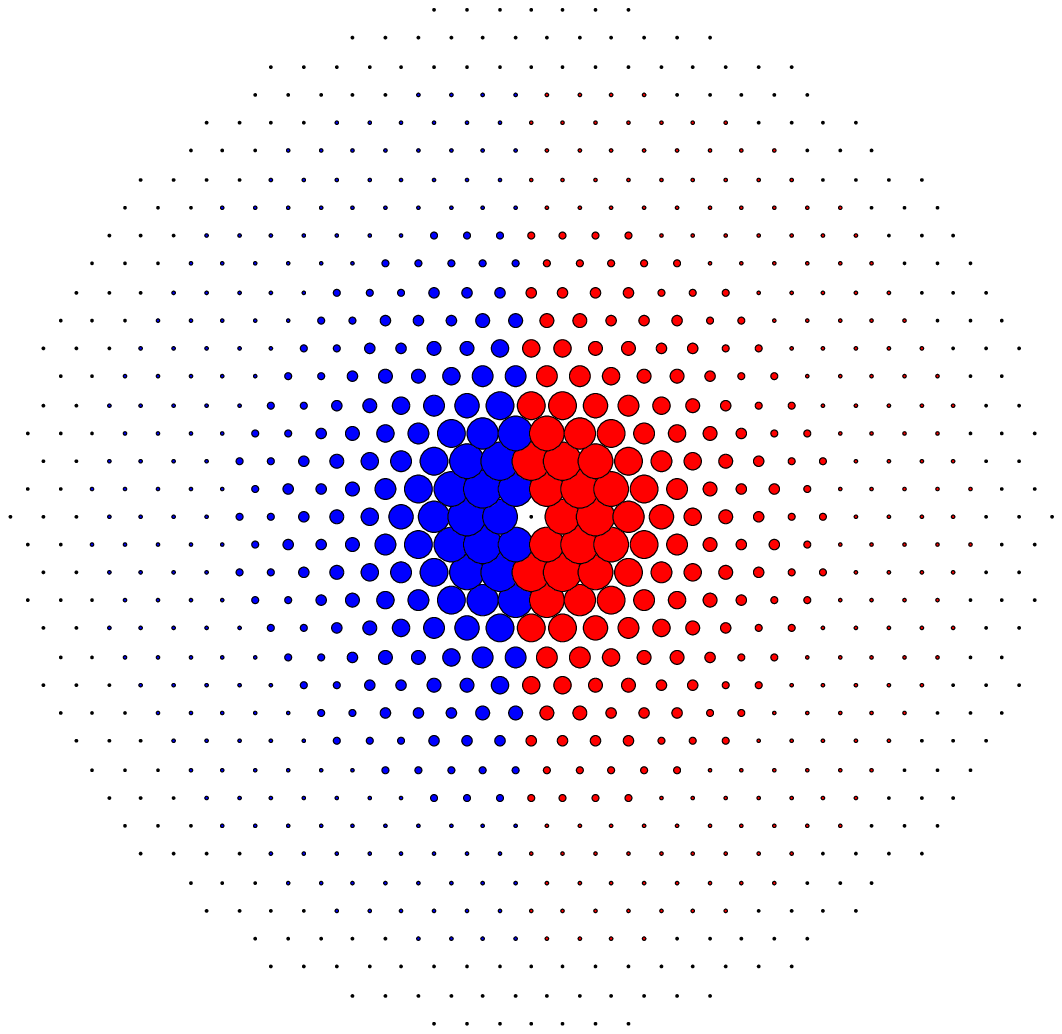}
\\
$m= - 1$                                                   &  $p= + 1$
\\
\hline
\end{tabular}
&
%
%
\begin{tabular}{|c|c|}
\hline
$\Delta_{+}$ & $\Delta_{-}$\\
\hline
\epsfxsize0.10\hsize\epsffile{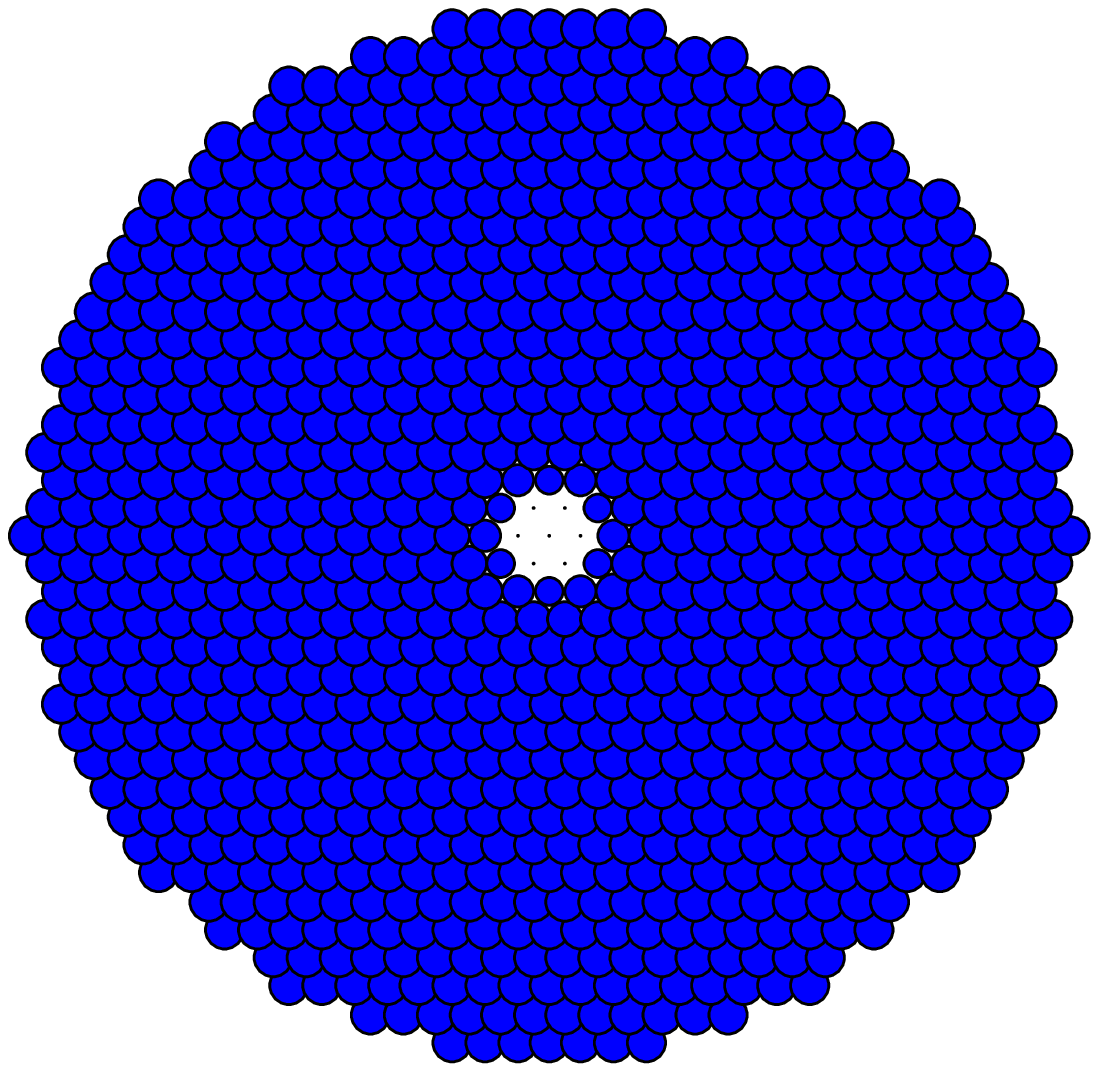} & \epsfxsize0.10\hsize\epsffile{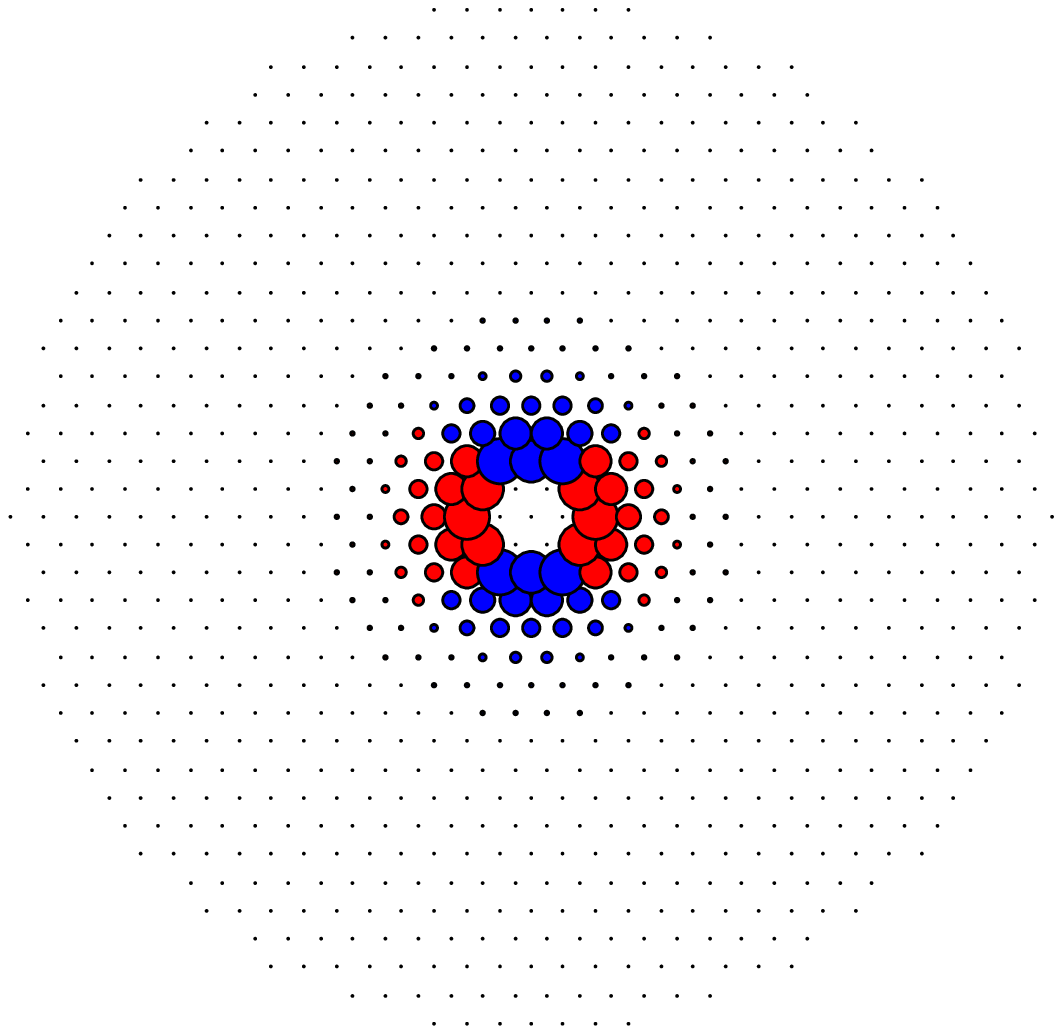} 
\\
$m= 0$                                                   &   $p= + 2$
\\
\hline
\end{tabular}
&
%
%
\begin{tabular}{|c|c|}
\hline
$\Delta_{+}$ & $\Delta_{-}$\\
\hline
\epsfxsize0.10\hsize\epsffile{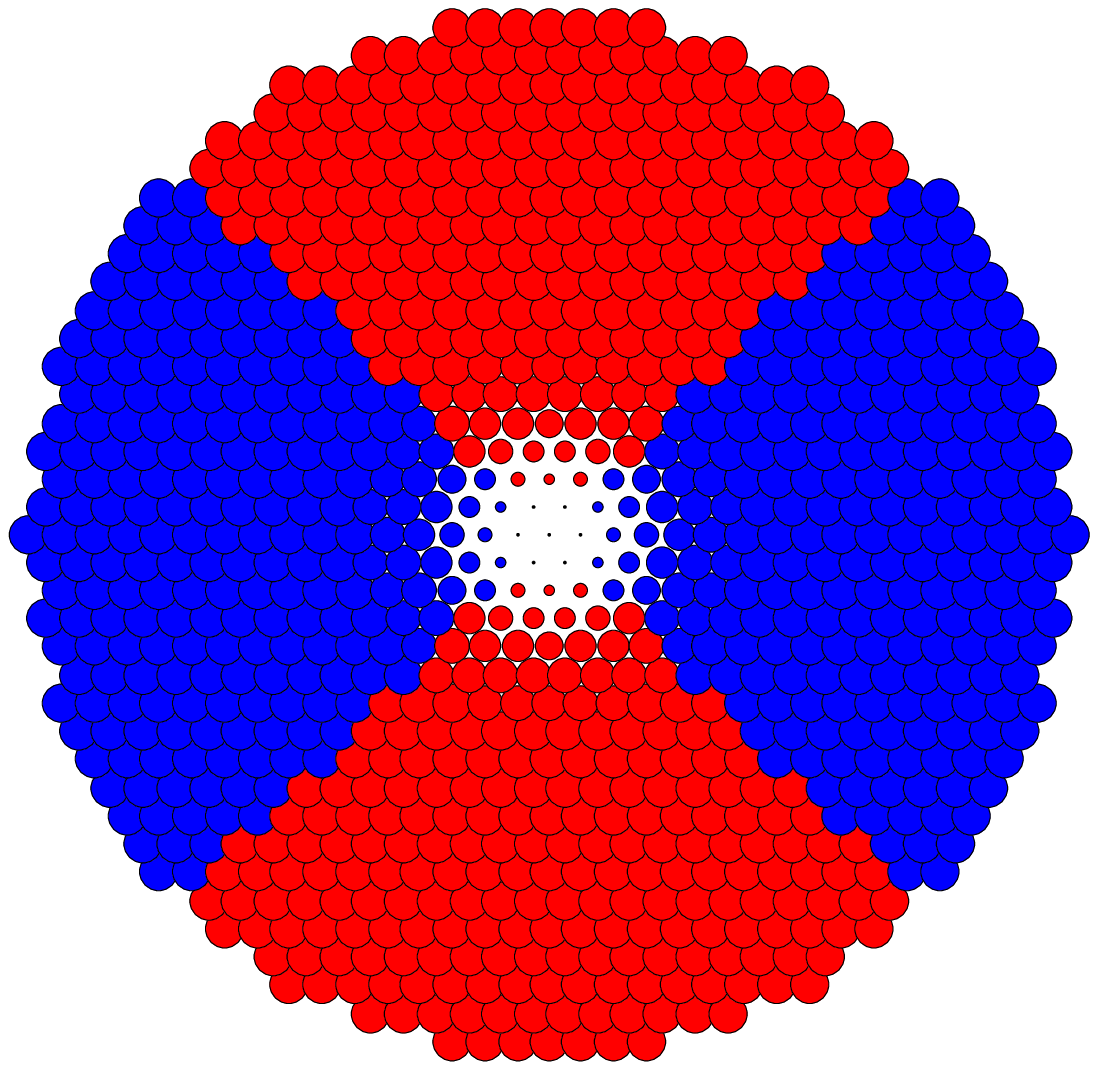} & \epsfxsize0.10\hsize\epsffile{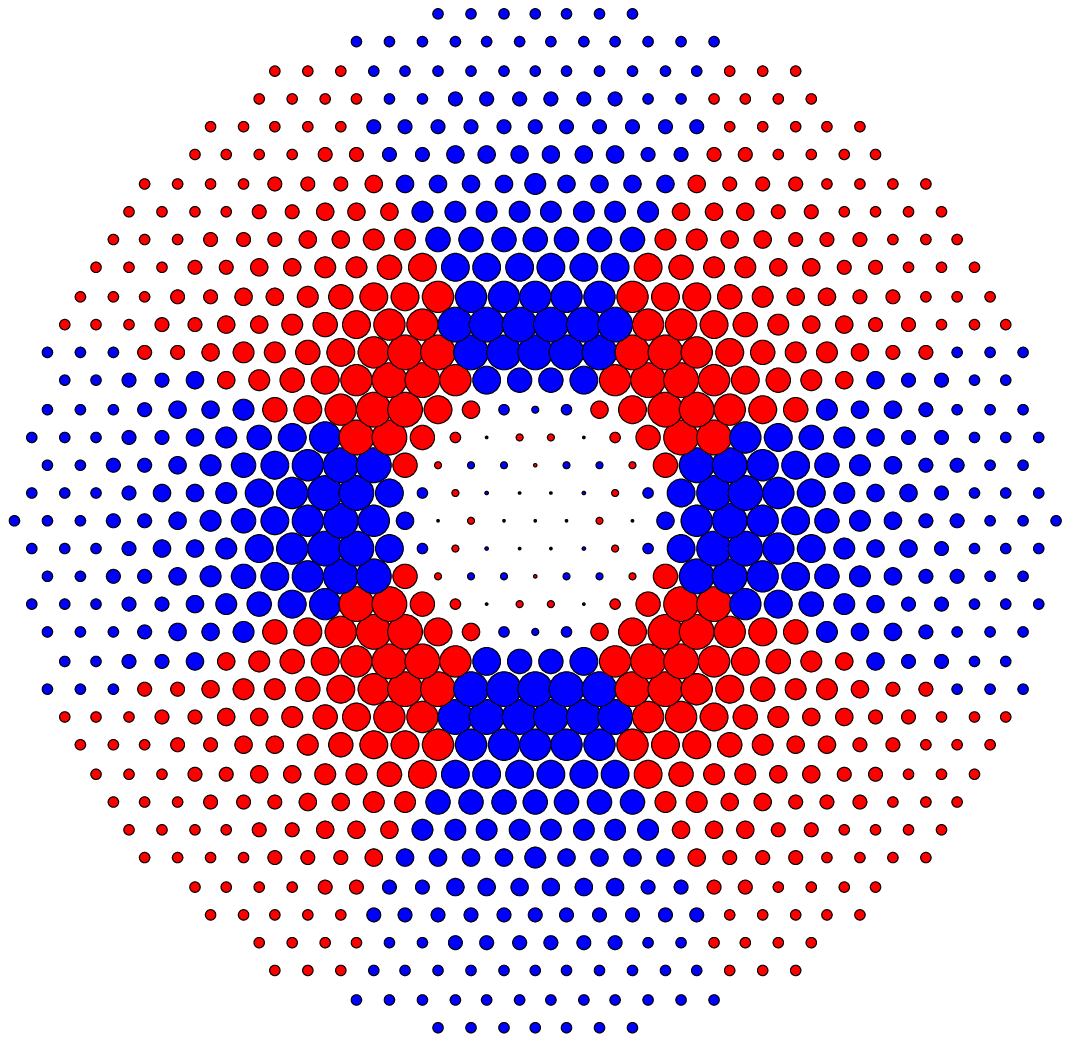} 
\\
$m= + 2$                                               & $p= + 4$ 
\\
\hline\hline
\epsfxsize0.10\hsize\epsffile{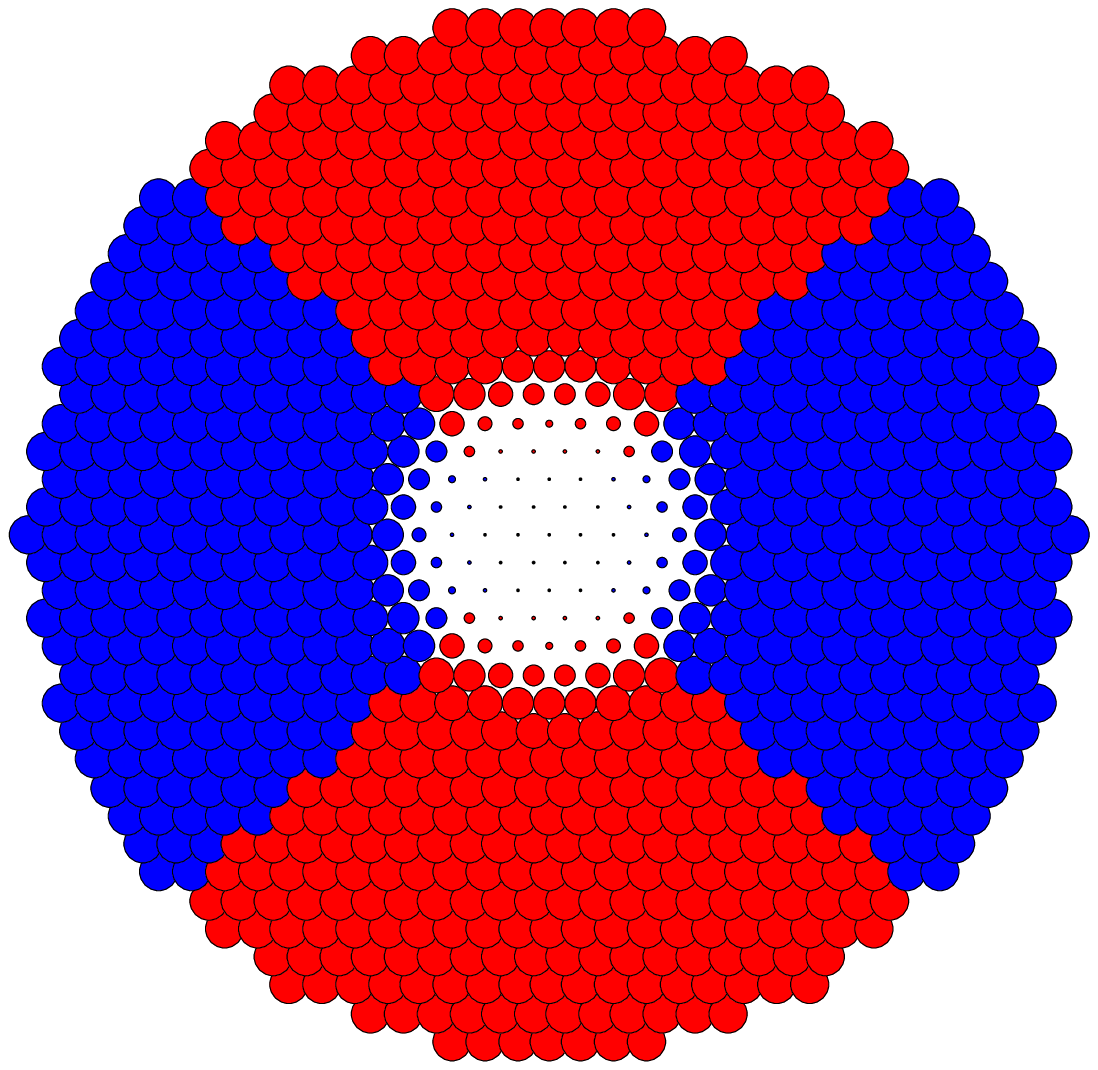} & \epsfxsize0.10\hsize\epsffile{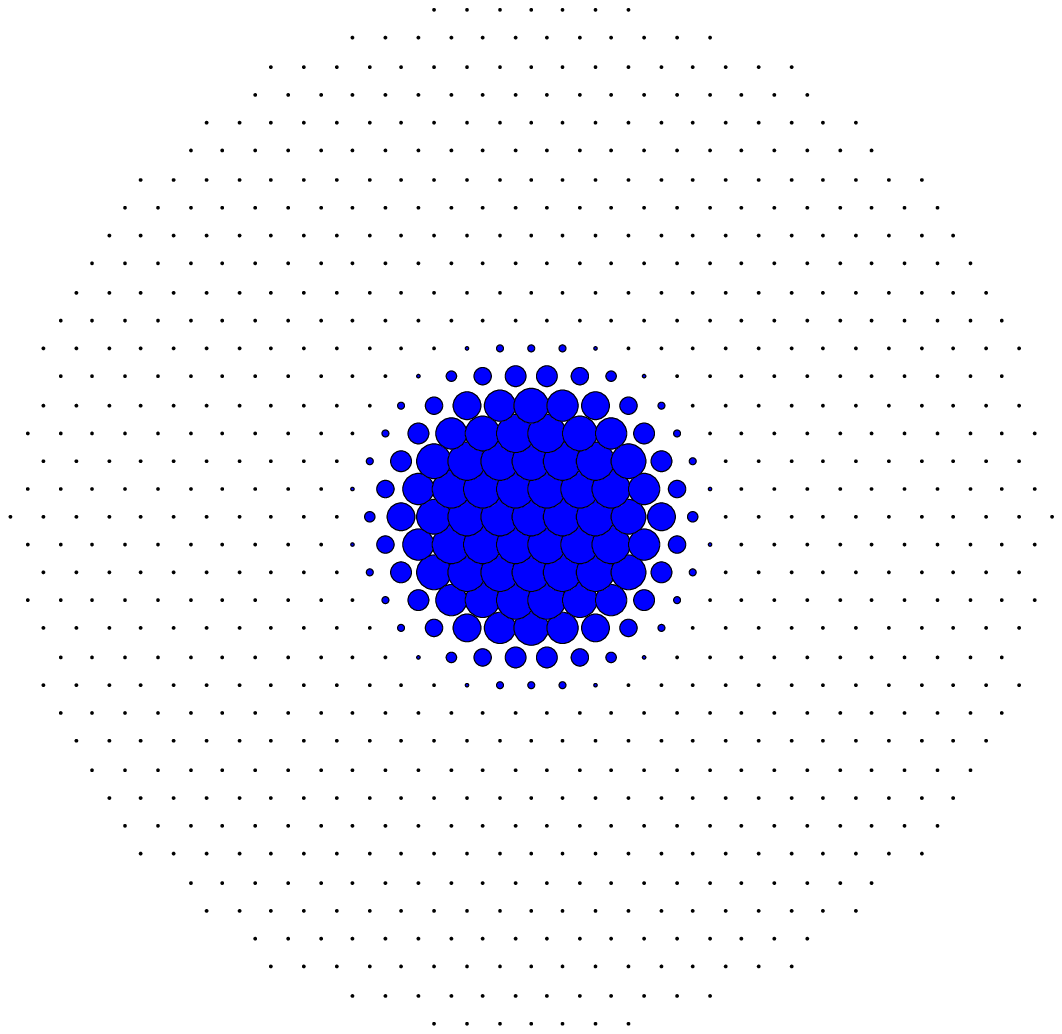}
\\
$m= - 2$                                                 &  $p= 0$
\\
\hline
\end{tabular}
\end{tabular}
}
\caption{Table of vortex states for the chiral ground state with $L_z^{\text{orb}}=+\hbar$. Shown are the dominant order parameter, $\Delta_{+}$, its global phase winding $m$, and the corresponding vortex core order parameter, $\Delta_{-}$, and its local phase winding $p$. The color code describes the phase winding - a change in color corresponds to a change of the sign in the real part of the corresponding order parameter amplitude. The order parameter with $m=0$ and $p=+2$ corresponds to a defect with no global phase winding, but with local phase winding near the defect.
}
\end{table}

Table \ref{table-winding_numbers} summarizes the lowest energy vortex states and their corresponding winding numbers. The figures show calculations of the dominant order parameter, $\Delta_{+}$, as well as the sub-dominant order parameter, $\Delta_{-}$, that develops in the core of the vortex. The color coding indicates the corresponding phase winding - a change in color corresponds to a change of the sign in the real part of the corresponding order parameter amplitude. The size of the circular symbols indicates the relative magnitude of the order parameter.  We discuss each of the vortex states in more detail below. 

Figures \ref{Fig1_1}, \ref{Fig1_2}, and \ref{Fig1_0} provide additional quantitative results for the order parameter structure, the current distribution and local density of states for the vortices listed in Table \ref{table-winding_numbers}. The magnitudes of the order parameter components, $\Delta_{+}$ and $\Delta_{-}$, are shown in column 3 as a function of distance along the $x$ axis through the vortex center.  The local density of states, ${\cal N}(\vp_f,\vR;\epsilon)$ is shown for a family of trajectories parallel to the $x$ axis with impact parameters $y$ separated by $\Delta y = 0.78\xi_0$.  Measurements of the local density of states using STM spectroscopy, which has been successfully applied to vortex studies in layered superconductors \cite{hes89f} including the high-$T_c$ cuprates \cite{hof02}, could provide valuable information on the vortex structure and pairing symmetry of \sro.

Our calculations were performed with impurity scattering included in Born limit for a mean free path of $\ell =10 \xi_0$, where $\xi_0=\hbar v_f/2\pi k_B T_c$ is the coherence length. The Fermi surface parameters are assumed to be isotropic, and the temperature was chosen to be $T=0.2\, T_c$. For simplicity we also assumed the high-$\kappa $ limit, where the penetration depth is large compared to the coherence length. Our calculations of the vortex structure are appropriate to the low-field limit near $H_{c1}$ where vortices are well separated from each other, or for $H < H_{c1}$ in the case of the defect with $m=0$.  The order parameter, impurity self energy, and the equilibrium spectra and current densities were obtained self consistently using the Riccati formulation of the quasiclassical transport theory with impurity and pairing self-energies. 

\subsection{Two types of singly quantized vortices}

\begin{figure}[h] 
\begin{tabular}{c}
$\Delta_{+}$
\\
\epsfxsize0.15\hsize\epsffile{figs/op2+1.ps} 
\\
$m = + 1$ 
\end{tabular}
\hspace{0.75cm}
\begin{tabular}{c}
$\Delta_{-}$
\\
\epsfxsize0.15\hsize\epsffile{figs/op3+1.ps}
\\
$p = + 3$
\end{tabular}
\hspace{0.75cm} 
\begin{tabular}{c}
\\
\epsfxsize0.22\hsize\epsffile{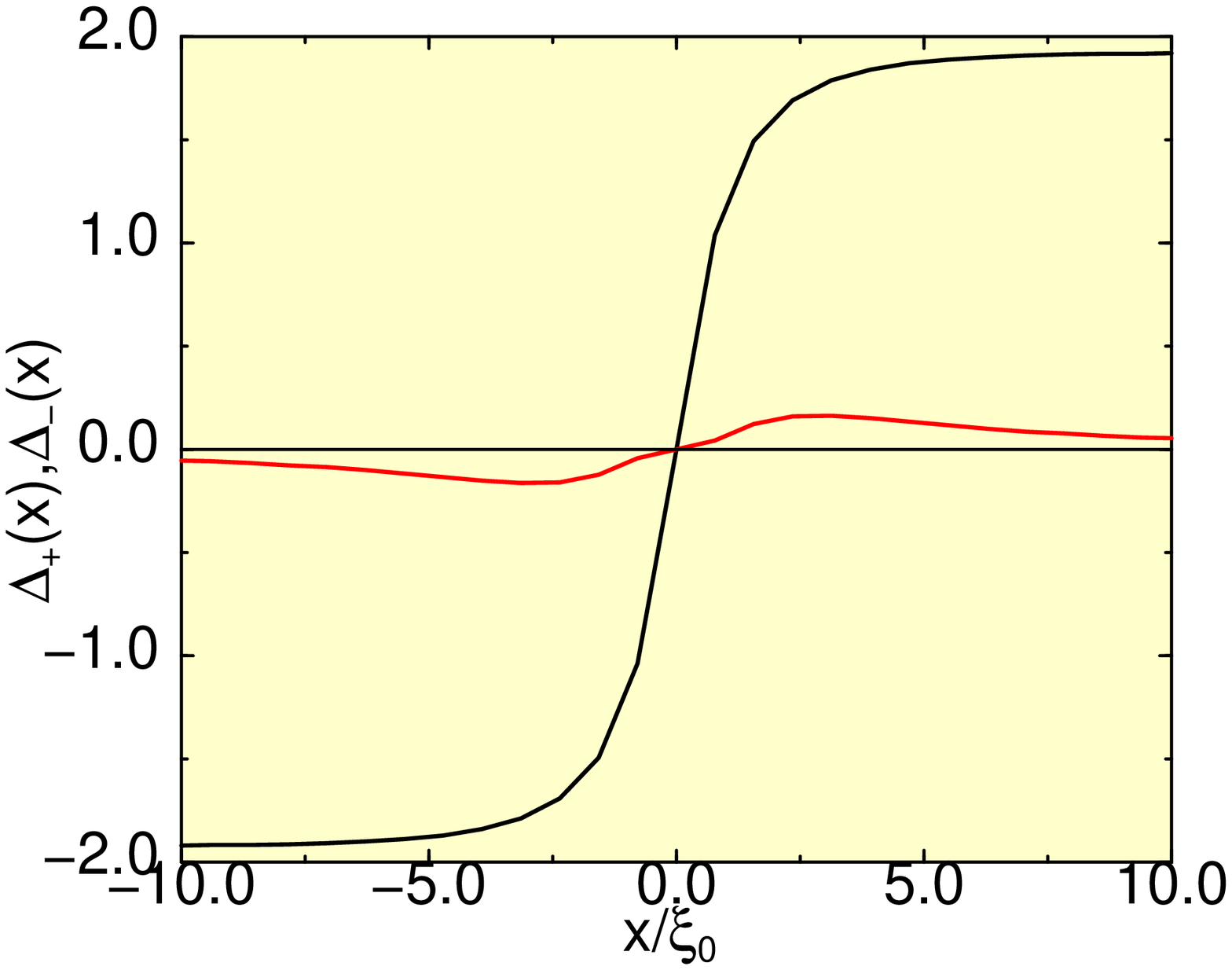} 
\\ 
\end{tabular}
\hspace{0.75cm} 
\begin{tabular}{c}
\\
\epsfxsize0.19\hsize\epsffile{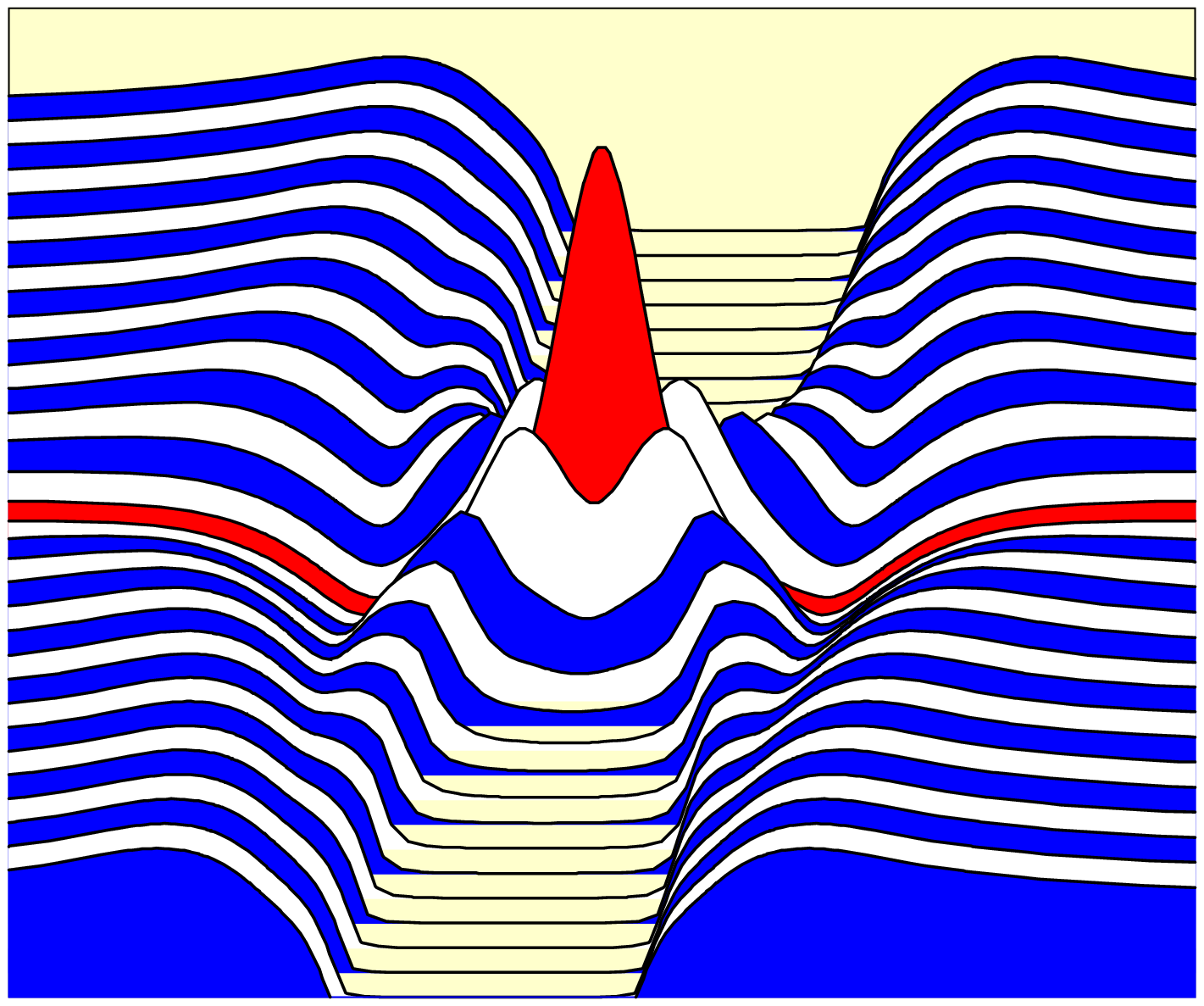} 
\\
\end{tabular}
\begin{tabular}{c}
\\
\epsfxsize0.15\hsize\epsffile{figs/op2-1.ps} 
\\
$m = - 1$ 
\end{tabular}
\hspace{0.75cm}
\begin{tabular}{c}
\\
\epsfxsize0.15\hsize\epsffile{figs/op3-1.ps}
\\
$p = + 1$
\end{tabular}
\hspace{0.75cm} 
\begin{tabular}{c}
\\
\epsfxsize0.22\hsize\epsffile{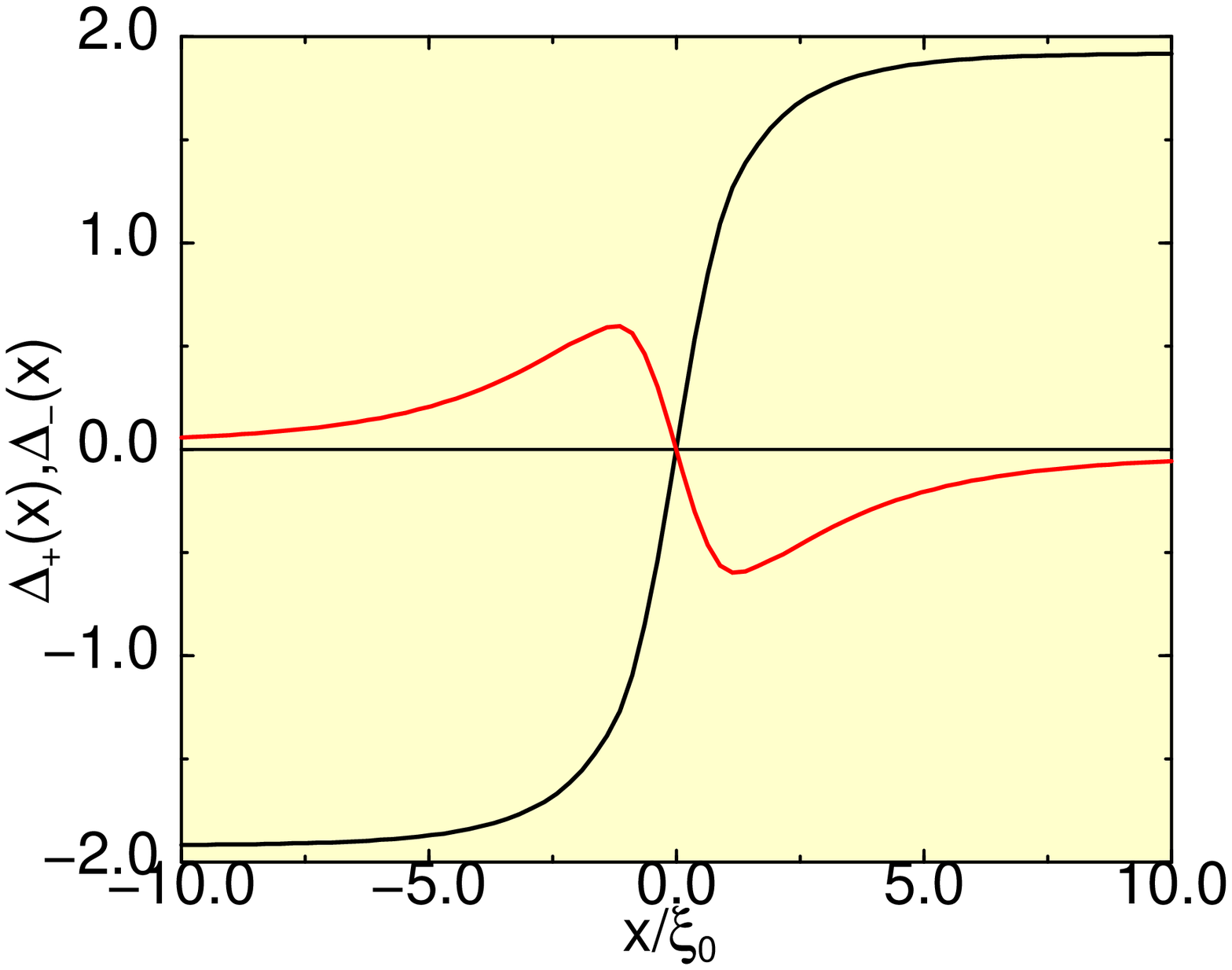} 
\\ 
\end{tabular}
\hspace{0.75cm} 
\begin{tabular}{c}
\\
\epsfxsize0.19\hsize\epsffile{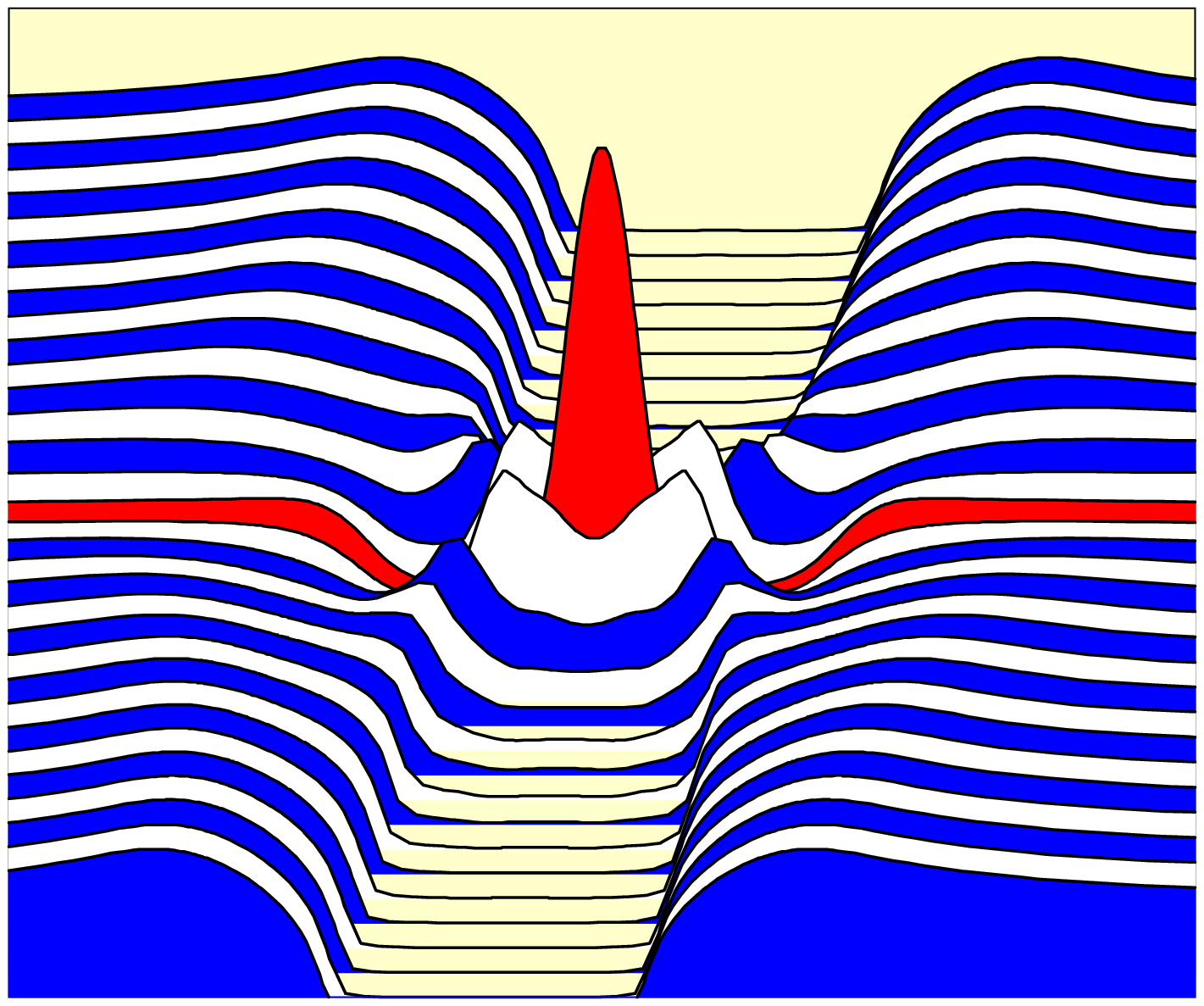} 
\\
\end{tabular}
\caption{ 
\label{Fig1_1}
Singly quantized vortices. Shown are the dominant order parameter $\Delta_{+}$ (first column) and the subdominant order parameter $\Delta_{-}$ (second column). The color coding is the same as in that of Table \ref{table-winding_numbers} and the size of the symbols reflects the magnitude of the order parameter. In the third column we show both order parameter components, $\Delta_{+}$ (black) and $\Delta_{-}$ (red), as a function of distance along a trajectory through the vortex center. In the fourth column the local density of states, $N(x=0,y;\epsilon)$, is shown as a function of energy and ``impact parameter'', $y$, from the the vortex center for $y=-12.5\xi_0$ to $y=12.5 \xi_0$ and a spacing of $\Delta y= 0.78\,\xi_0$. The red spectrum showing the zero-energy core state is evaluated at $x=0$, $y=0$.
}
\end{figure}

Broken time-reversal symmetry of the ground state leads to an interesting effect: vortices with opposite global phase winding numbers are \textsl{inequivalent} \cite{tok90}. The ground state pairs have internal orbital angular momentum, $\textbf{L}_{z}^{\text{orb}}$, which can be either parallel or antiparallel to the external field $\vH  = H\hat{\vz}$ that nucleates a vortex and fixes the sign of the global phase winding. For $\vH  || \textbf{L}_z = +\hbar\hat{\vz}$ the $m=1,p=3$ vortex is realized, whereas for the antiparallel case, $-\vH  || \textbf{L}_z = +\hbar\hat{\vz}$ or $\vH  || \textbf{L}_z = -\hbar\hat{\vz}$, the $m=-1,p=+1$ vortex is nucleated. The structures of both vortices are shown in Fig. \ref{Fig1_1}. Both vortices exhibit the large zero-energy Andreev bound state at the vortex center that is characteristic of odd winding number \cite{rai96}. Note however the difference in magnitude of the induced components for $m=\pm 1$. The relative orientation of the field, $\vH $, and the spontaneously chosen direction for the internal orbital angular momentum of the pairs in the ground state, $\textbf{L}_z^{\text{orb}} = \pm\hbar\hat{\vz}$, determines which type of singly quantized vortex is nucleated from the Meissner state at the lower critical field. The $m=+1$ vortex has a local phase winding of $p=+3$, and thus different core energy than the simpler $m=-1$ vortex with $p=+1$. 
As originally noted by Tokuyasu et al. \cite{tok90}, the difference in free energy for these two vortices leads to a splitting of the lower critical field, $H_{c_{1+}}\ne H_{c_{1-}}$, for fields parallel (+) or antiparallel (-) to the internal orbital angular momentum of the pairs, $\textbf{L}_z^{\text{orb}}$ \cite{tok90}. Observation of this asymmetry would provide a direct signature of broken chirality in the ground state.

Ginzburg-Landau (GL) calculations by Tokoyasu et al. \cite{tok90} also predicted the possibility of \textsl{spontaneously broken axial symmetry} by the cores of singly quantized vortices with $m=\pm 1$. In the GL theory whether or not the axial symmetry is spontaneously broken is determined by competition between the suppression of the condensation energy by the global phase winding and the internal Josephson phase-locking energy between the two time-reversed order parameter amplitudes. If the GL coefficient, $\beta_{2}$, that determines the coupling energy, $F_{\text{c}} = 4\beta_{2} |\Delta_{+}|^2\,|\Delta_{-}|^2$, is sufficiently weak compared with the condensation terms, $\beta_1\,\left[|\Delta_{+}|^4+|\Delta_{-}|^4\right]$, i.e. $0 < \beta_{2} \le \tinyonefourth \beta_{1}$ then axial symmetry is spontaneously broken by the vortex core. Our results for singly quantized vortices, which do not break axial symmetry at any temperature, also agree with those of Ref. \cite{tok90} since our theory reduces to the GL equations with the weak-coupling value of $\beta_{2}/\beta_{1} = \tinyonehalf$ in the limit $T\approx T_c$.

Other authors have also considered differences between singly quantized vortices with opposite phase windings relative to the chirality of the ground state. 
The authors of Ref. \cite{ich02} report results for the field dependence of the vortex structure in chiral p-wave superconductors. They argue that H$_{c_{2}}$ differs for states with opposite chirality, i.e. $L_{z}^{\text{orb}} = \pm \hbar$. However, this is incorrect. 
The upper critical field represents a second-order transition between the normal state and the superconducting vortex state. Since chiral symmetry is {\sl unbroken} in the normal state, the second-order instability field, $H_{c_{2}}$, is necessarily the same for $\vH  || +\hat{\vz}$ or $\vH  || -\hat\vz$. Furthermore, at  H$_{c_{2}}$ \ul{both} chiral components of the 2D representation are nucleated. As the field is reduced, the vortex density decreases and eventually a single chiral domain may remain, at least in the idealized limit of a perfect crystal with no vortex pinning. The chirality of the resulting ground state in the limit $H\rightarrow 0$ will be determined by the direction of $\vH $, but H$_{c_{2}}$ is the same for either field orientation.
The situation is completely different for $H_{c_{1}}$. In this case a vortex enters the superconductor from the Meissner state which spontaneously breaks chiral symmetry. The relative sign of the phase winding of the vortex and the chirality of the pre-existing Meissner phase lead to different vortex core states nucleating depending on whether $\vH  || +\hat\vz$ or $\vH  || -\hat\vz$, and a corresponding asymmetry in $H_{c_{1}}$.

Recently, Yokoyama et al.\cite{yok08} proposed that broken chiral symmetry should be observable as an asymmetry in the surface density of states (SDOS) at the Fermi level when a vortex is situated within a coherence length of the surface of a chiral p-wave superconductor. These authors predict a large anomaly in the SDOS depending on the sign of the chirality based on a non-self-consistent model for the vortex core which omits the internal structure of the vortex cores of chiral superconductors described above.

Similarly, the authors of Ref. \cite{kat02} report a dramatic difference in the broadening of the energy levels in the vortex core for the $m=+1$ and $m=-1$ vortices.\footnote{The $m=+1$ vortex in the chiral ground state with $\textbf{L}_z^{\text{orb}} = +\hbar\hat{\vz}$ corresponds to the $L_z = +2$ vortex of Ref. \cite{kat02}, while the $m=-1$ vortex in the same chiral ground state corresponds to their $L_z=0$ vortex.} Their analysis is based on an assumed form for the vortex amplitude that does not include the induced component of the order parameter in the core with time-reversed chirality. Our results, which are based on fully self-consistent solutions for the order parameter, single-particle self-energy and spectral density, show only a minor reduction in the spectral width of the $m=-1$ core state compared to that for the $m=+1$ core state.
This is consistent with earlier calculations by Hess et al. \cite{hes96,hes96a}, who found a small asymmetry in the dispersion of the Andreev bound states for self-consistent solutions for singly quantized vortices in the chiral ground state.

The discrete symmetry of the point group is not expected to significantly change the structure of the $m=-1,p=1$ vortex since the next relevant higher order components are $m=-1,p=-3,+5\,\ldots$. However, the $m=+1,p=+3$ vortex allows a higher-order subdominant component with $p=-1$, which may have a significant amplitude since it will be localized relatively close to the vortex center. The corrections to the vortex core resulting from higher order harmonics to the basis functions allowed by the discrete point group symmetry will reduce the difference in free energy between the two types of singly quantized vortices.
Takigawa et al. \cite{tak02} have considered the extreme anisotropic limit for the structure of singly quantized vortices and the local DOS by solving the Bogoliubov equations numerically on a lattice with nearest neighbor (NN)  hopping as well as local NN pairing interactions.

\subsection{Doubly quantized vortices}

\begin{figure}[t]
\begin{tabular}{c}
$\Delta_{+}$
\\
\epsfxsize0.15\hsize\epsffile{figs/op2+2.ps} 
\\
$m = + 2$ 
\end{tabular}
\hspace{0.75cm}
\begin{tabular}{c}
$\Delta_{-}$
\\
\epsfxsize0.15\hsize\epsffile{figs/op3+2.ps}
\\
$p = + 4$
\end{tabular}
\hspace{0.75cm} 
\begin{tabular}{c}
\\
\epsfxsize0.22\hsize\epsffile{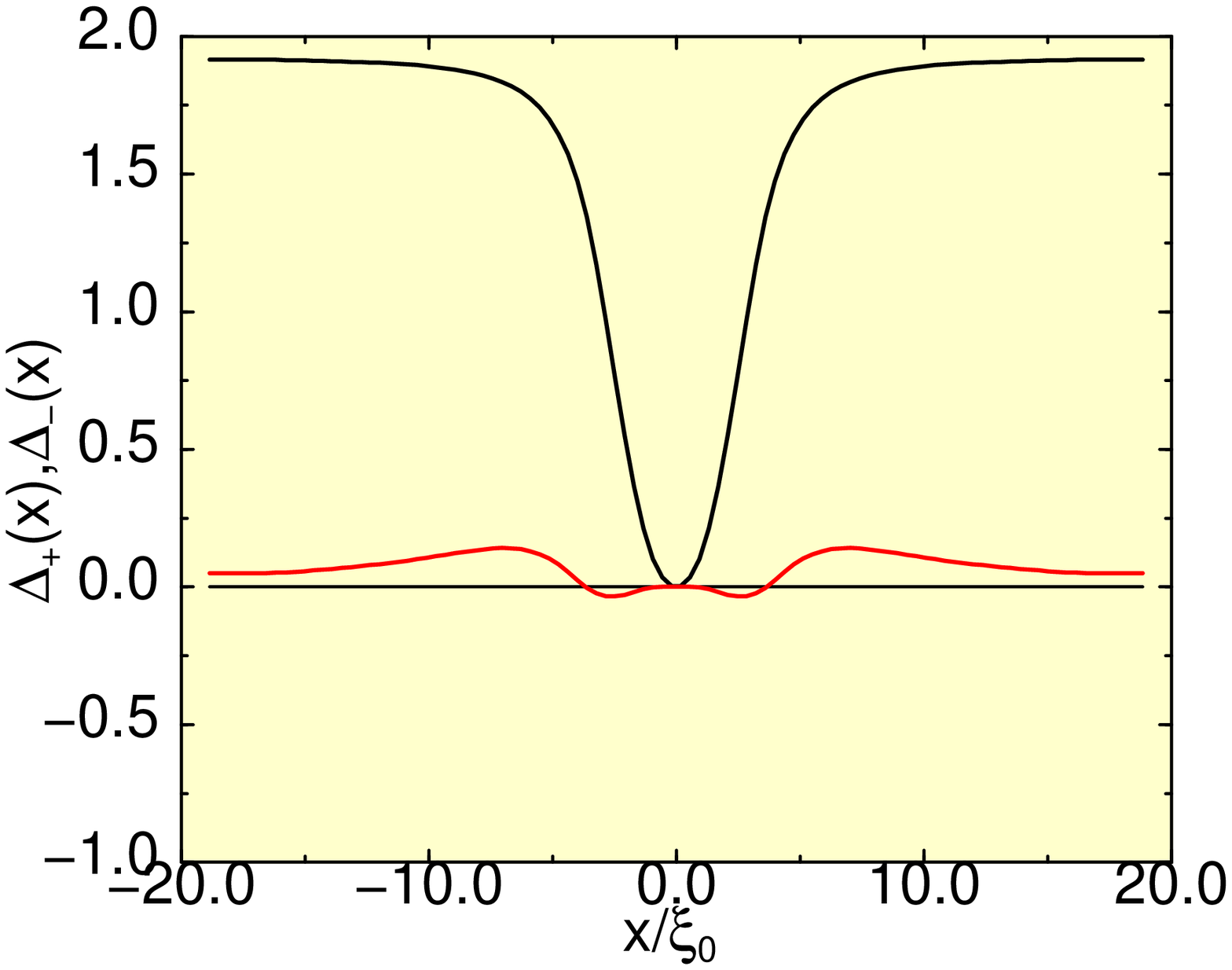} 
\\ 
\end{tabular}
\hspace{0.75cm} 
\begin{tabular}{c}
\\
\epsfxsize0.19\hsize\epsffile{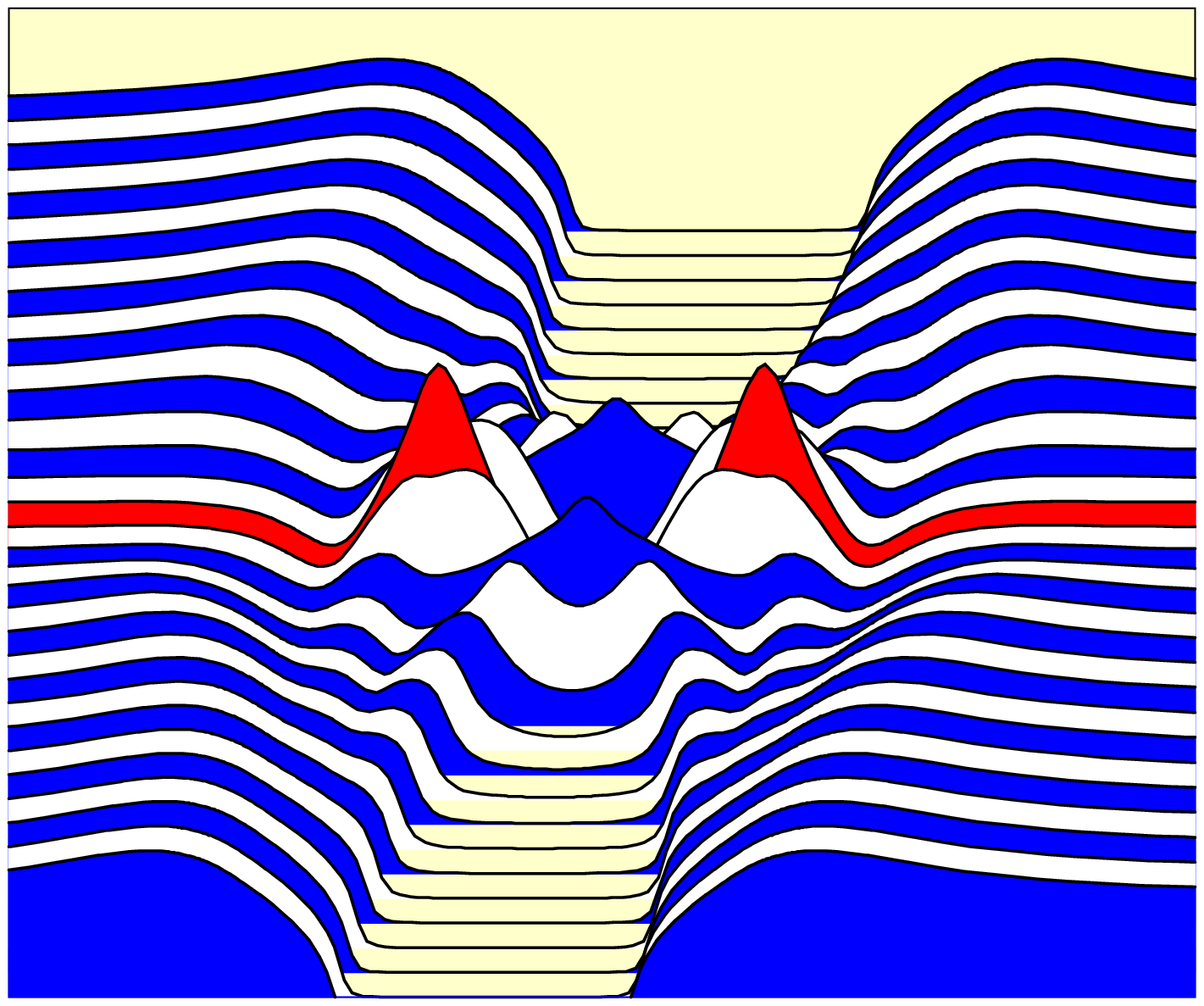} 
\\
\end{tabular}
\begin{tabular}{c}
\\
\epsfxsize0.15\hsize\epsffile{figs/op2-2.ps} 
\\
$m = - 2$ 
\end{tabular}
\hspace{0.75cm}
\begin{tabular}{c}
\\
\epsfxsize0.15\hsize\epsffile{figs/op3-2.ps}
\\
$p =  0$
\end{tabular}
\hspace{0.75cm} 
\begin{tabular}{c}
\\
\epsfxsize0.22\hsize\epsffile{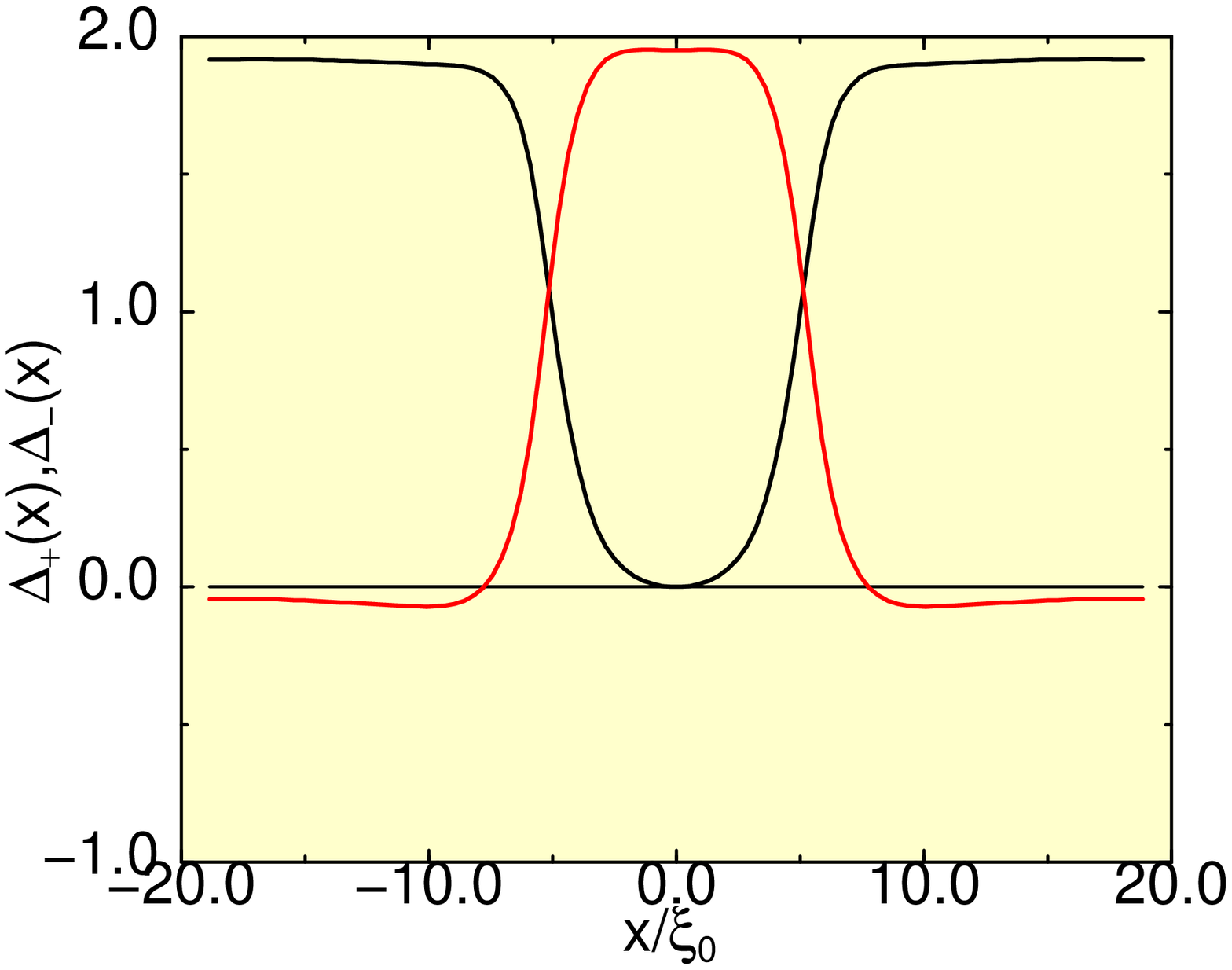} 
\\ 
\end{tabular}
\hspace{0.75cm} 
\begin{tabular}{c}
\\
\epsfxsize0.19\hsize\epsffile{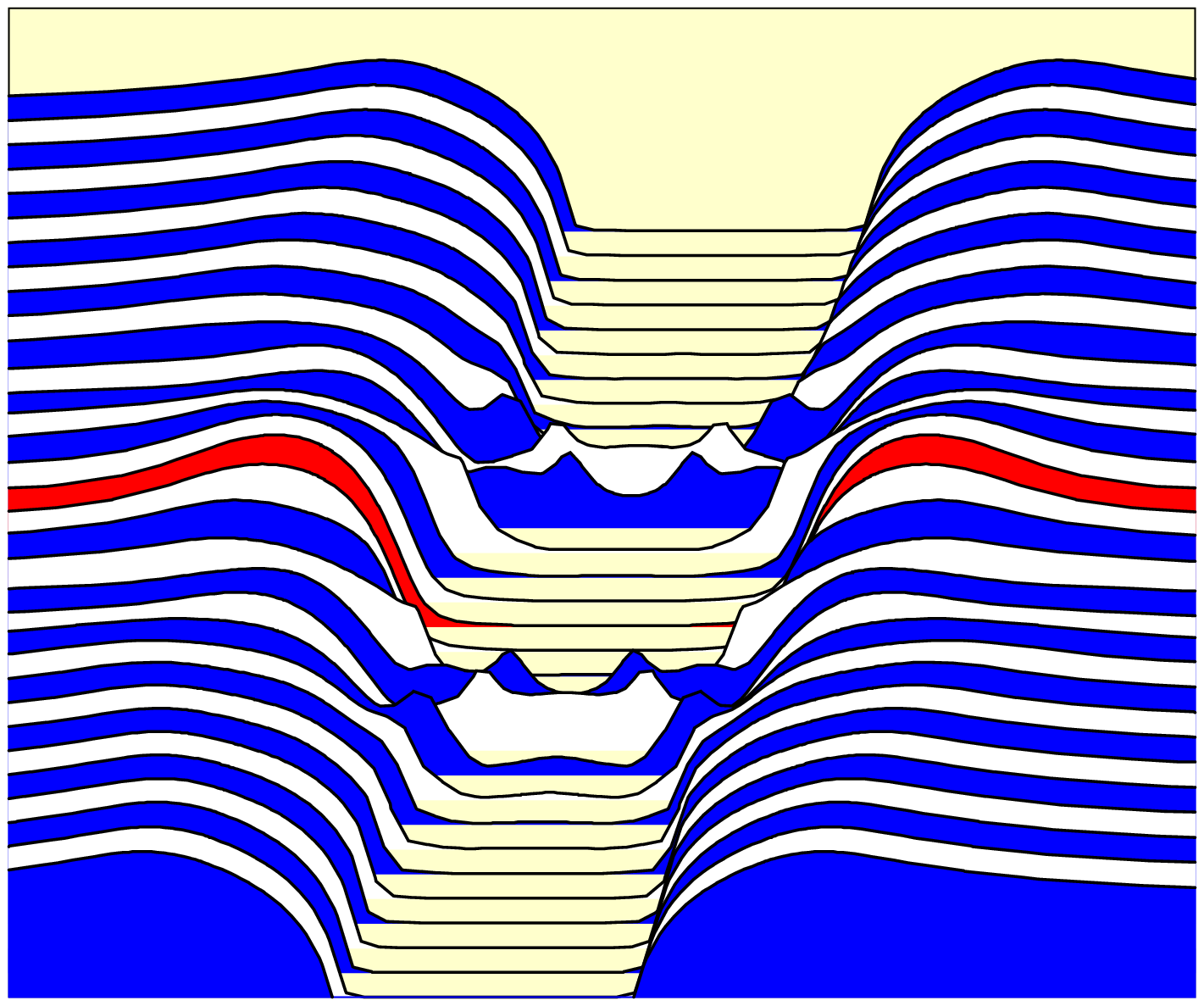} 
\\
\end{tabular}
\caption{ 
\label{Fig1_2}
Doubly quantized vortices. The columns and notation are the same as that of Fig. \ref{Fig1_1}. Note (i) the absence of a zero-energy state at the vortex center, (ii) the small sub-dominant amplitude, $\Delta_{-}$, for the $m=+2$, $p=+4$ vortex, (iii) the large sub-dominant amplitude that fills the core of the $m=-2$, $p=0$ vortex (iv) and the correspondingly low density of sub-gap excitations in the vortex core.
}
\end{figure}

\begin{figure}[t]
\begin{tabular}{c}
\epsfxsize0.8\hsize\epsffile{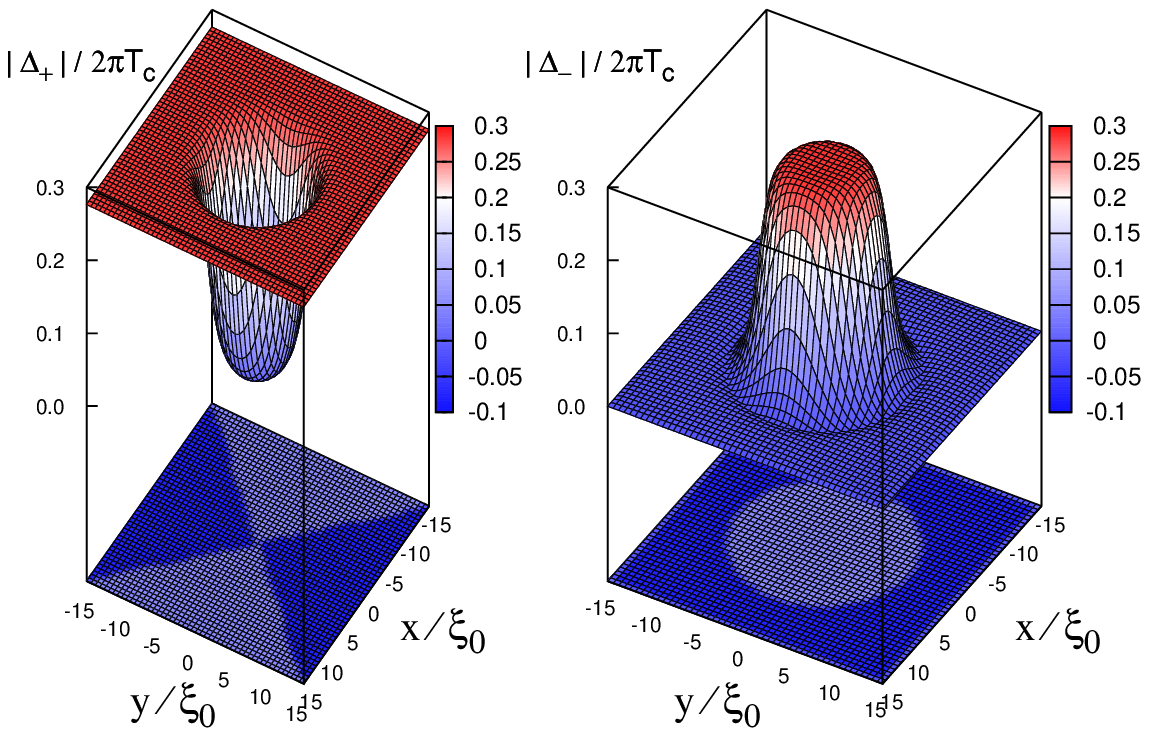} 
\\
\epsfxsize0.35\hsize\rotate[r]{\epsffile{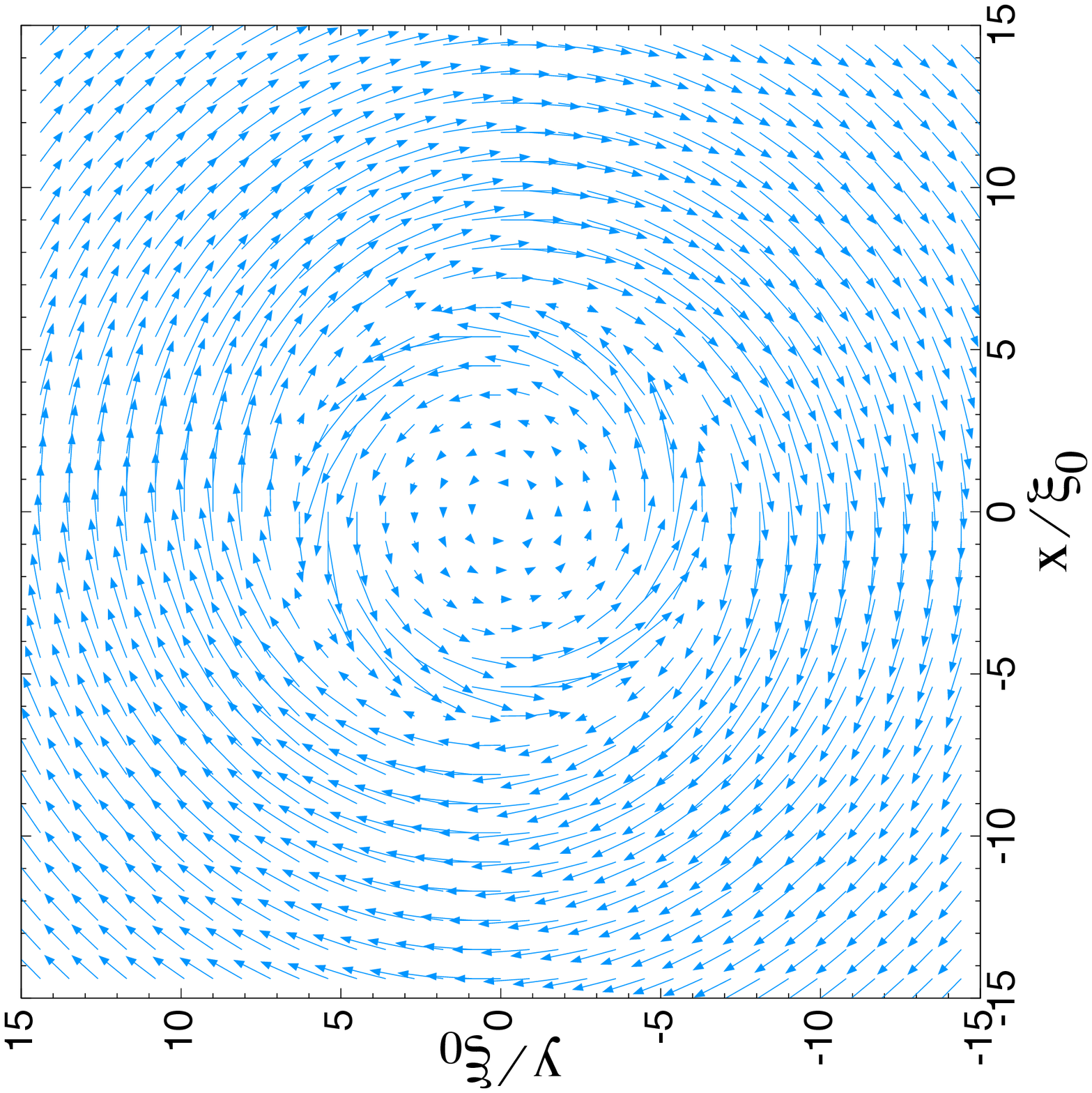}}
\end{tabular}
\caption{ 
\label{Fig2_-20}
Axially symmetric double quantum vortex with $m=-2$ and $p=0$. Top-left: the dominant amplitude $|\Delta_{+}(\vR)|/2\pi k_B T_c$ is strongly suppressed in the core. The projection in the plane is a phase plot of $\sgn(\Re[\Delta_{+}(\vR)\exp(im\phi)])$. Top-right: the amplitude, $|\Delta_{-}(\vR)|/2\pi k_B T_c$, develops to its equilibrium value and fills the vortex core, while the phase plot of $\sgn(\Re[\Delta_{-}(\vR)\exp(ip\phi)])$ shows a radial $\pi$ phase change indicative of a domain wall separating the time-reversed phase in the core. Bottom: the domain wall separates an axially symmetric flow in the core that is counter-circulating relative to the current outside the core.
}\end{figure}

As in the case of singly quantized vortices, there are two in-equivalent doubly quantized vortices: (i) $m=2$, $p=4$ and (ii) $m=-2$, $p=0$ as shown in Fig. \ref{Fig1_2}. The latter case describes a doubly quantized vortex with a nearly homogeneous core amplitude, i.e. a "coreless vortex". Since there is no local phase winding associated with the subdominant amplitude, it develops to the asymptotic value for the dominant order parameter component as indicated in the third column of Fig. \ref{Fig1_2}. 
This is similar to a domain wall separating the two degenerate order parameter components, except that the subdominant component is restricted to the core (see Fig. \ref{Fig2_-20}). The large subdominant component which ``fills in'' the core of the doubly quantized vortex substantially lowers the free energy for the $m=-2$, $p=0$ vortex by recovering nearly all the lost condensation energy from the core of the dominant component. 
This low core energy is reflected in the local density of states shown in the lower panel of column 4 in Fig. \ref{Fig1_2}; there are very few low-energy, sub-gap excitations associated with pair-breaking in the vortex core.
The large reduction in the core energy of the doubly quantized vortex compared to that of the singly quantized vortex allows for stable lattices of doubly quantized vortices at sufficiently high magnetic fields. The possibility of stable lattices of doubly quantized vortices in superconductors with broken chiral symmetry was investigated within GL theory. It was shown that doubly quantized vortices can be stabilized compared to the singly quantized vortices over a wide range of fields below $H_{c2}$ \cite{tok90a}. The axial symmetry of the current distribution (shown in the lower panel of Fig. \ref{Fig2_-20}) leads to isotropic interactions between these doubly quantized vortices, and thus a hexagonal lattice structure.

\subsubsection*{Broken Axial Symmetry in the Vortex Core}

\begin{figure}[t]
\begin{tabular}{c}
\epsfxsize0.8\hsize\epsffile{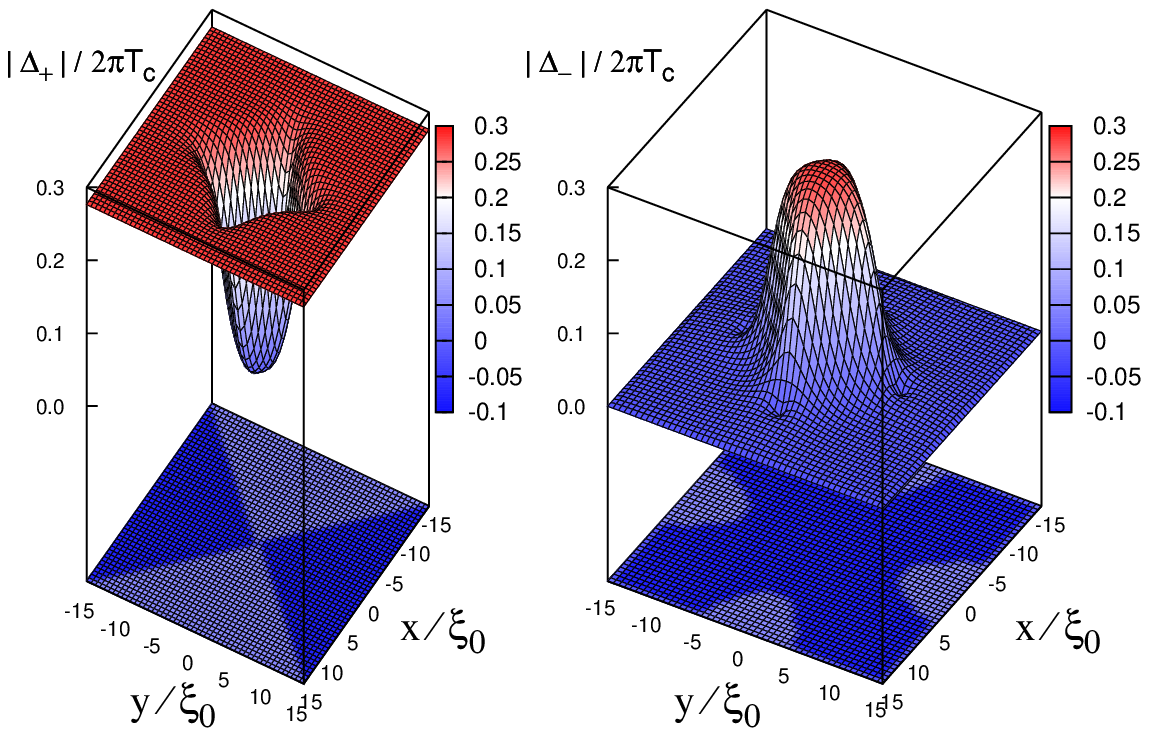} 
\\
\epsfxsize0.35\hsize\rotate[r]{\epsffile{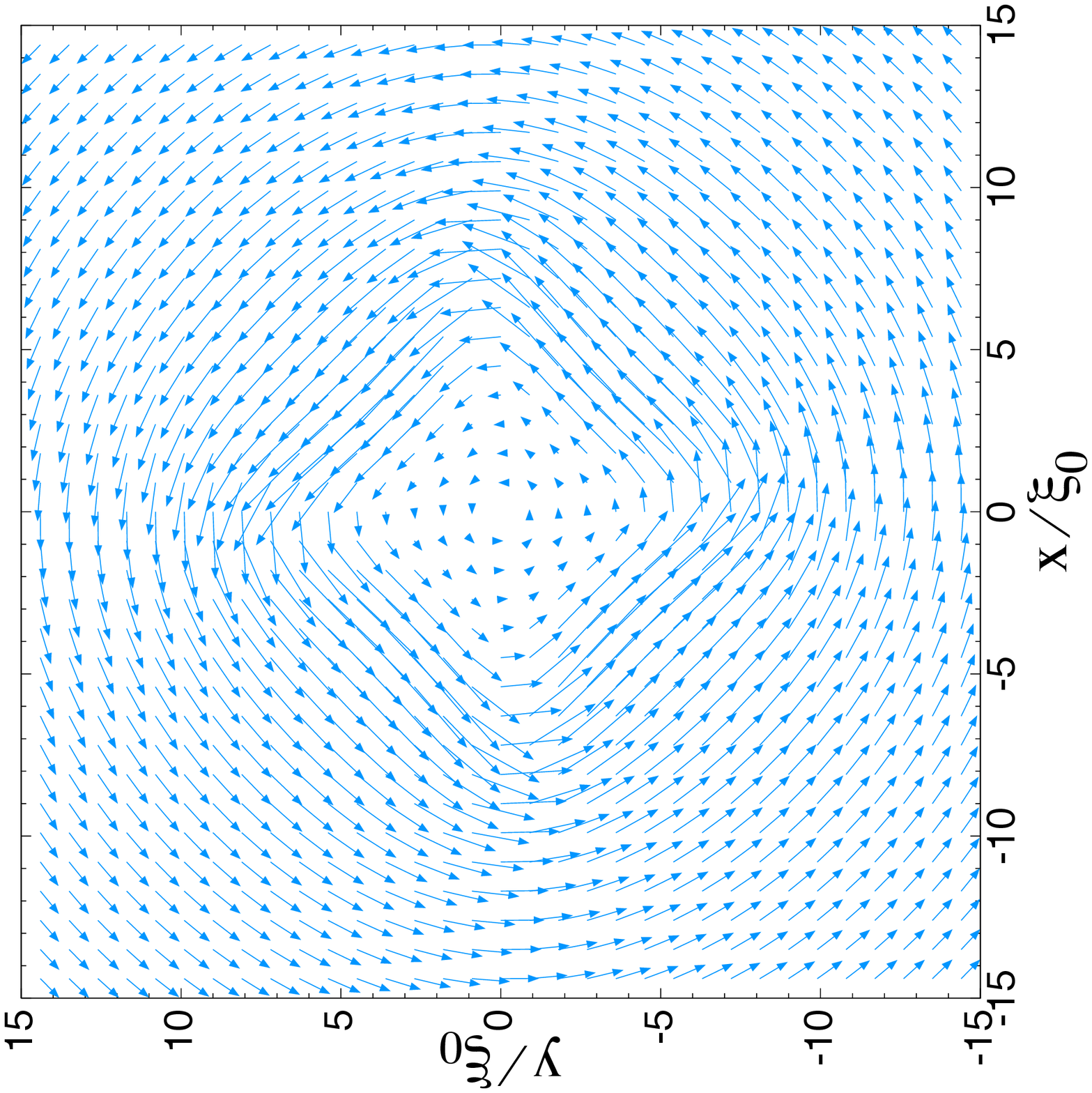}}
\end{tabular}
\caption{ 
\label{Fig2_24}
Spontaneously broken axial symmetry in the core of the double quantum vortex with $m=+2$ and $p=+4$. Top-right: The $p=+4$ vortex in the $\Delta_{-}$ dissociates into four singly quantized vortices situated as satellites at the corners of a square on the core boundary (note also the phase plot in the projection), allowing the amplitude $|\Delta_{-}|$ to grow substantially in the core. The satellite vortices of  $\Delta_{-}$ lead to 4-fold anisotropy in the dominant amplitude $|\Delta_{+}|$ (top-left panel), as well as a current distribution (bottom panel) which exhibits the four-fold anisotropy induced by the four $p=+1$ core vortices. Note that there are no counter-circulating currents once axial symmetry is broken.
}\end{figure}

The calculation of the structure of the doubly quantized vortex with $m=+2$ and $p=+4$ shown in Fig. \ref{Fig1_2} is for a relatively small area grid. The result shows an axially symmetric structure in which the local phase winding of $p=+4$ is concentrated as a multiply-quantized axially symmetric vortex. However, this structure is unstable. The stable solution for this vortex spontaneously breaks axial symmetry, with the $p=+4$ vortex in the time-reversed phase dissociating into \textsl{four} singly quantized vortices arranged around the edge of the time-reversed core amplitude as shown in the top-right panel of Fig. \ref{Fig2_24}. The breaking of $C_{\infty}$ to $C_{4}$ symmetry is energetically preferred because dissociation of the $p=+4$ vortex allows the time-reversed amplitude in the core to recover to the bulk value and restore most of the lost condensation energy of the dominant phase. This reduces the energy splitting between the two types of doubly quantized vortices and implies that the region of stability of doubly quantized vortices will be comparable for either field orientation, $\vH  || \textbf{L}_z = +\hbar\hat{\vz}$ or $-\vH  || \textbf{L}_z = +\hbar\hat{\vz}$. However, the axial anisotropy of the current distribution of the doubly quantized vortex (lower panel of Fig. \ref{Fig2_24}) is expected to lead to a square vortex lattice of doubly quantized vortices for $\vH  || \textbf{L}_z = +\hbar\hat{\vz}$, compared to a triangular lattice for $-\vH  || \textbf{L}_z = +\hbar\hat{\vz}$.


Finally, let us consider the effects on vortex structure from additional order parameter components which are allowed by discrete point symmetry of Fermi surface. For the axially symmetric double quantum vortex with $m=-2$ and $p=0$ the next order harmonics are those with $p=\pm 4$. The magnitudes will be very small because both amplitudes, $\Delta_{+}$ and $\Delta_{-}$, are large and there is no energetic advantage to inducing these components. Thus, we can safely neglect these higher harmonics.
The same reasoning applies for the vortex with $m=+2$ and $p=+4$. Both the dominant and core amplitudes are large and suppress the higher harmonics, even the higher harmonic with $n=-4$ (i.e. $p=0$). This would not be the case if the axial symmetry were not spontaneously broken.

\subsection{Structure of inhomogeneities in the Meissner phase}

In addition to novel vortex states, broken chirality leads to defect structures that carry current, but have no global phase winding. The example of a cylindrical normal metallic inclusion is included in Table \ref{table-winding_numbers}. This defect has no global phase winding, i.e. $m=0$, but necessarily has a local phase winding of $p=+2$ for the induced order parameter. Thus, in contrast to vortices that are stabilized by a field, this defect can exist in the Meissner phase. Such defects can be engineered, or as in the case of \sro,  Ru ions may precipitate and form metallic inclusions embedded in the superconducting host material.

The ground state of a chiral, p-wave superconductor will be either the $\hat{\vp}_x+i\hat{\vp}_y$ state, or the $\hat{\vp}_x-i\hat{\vp}_y$ state, although the actual zero-field superconducting phase may include a number of domains of different time-reversed states separated by domain walls. Here we consider a normal metallic inclusion of radius $r_{\text{incl}} = 0.78\xi_0$ embedded in the homogeneous $\hat{\vp}_x+i\hat{\vp}_y$ ground state, or at least several coherence lengths away from a domain wall.
We assume a simple model for a metallic inclusion in which electrons and holes are perfectly transmitted across the (NS) interface between the inclusion and the superconducting material when the latter is in its normal state. We also assume there is no pairing interaction within the metallic inclusion. Thus, a mean field order parameter does not develop inside the inclusion, however pairing correlations and associated spontaneous currents do develop within the inclusion as a result of the proximity effect and the high transmission of the NS interface. 
More detailed models, including specific models of Ru inclusions, which incorporate the finite reflectivity of the NS interface, as well as pairing attraction within the metallic inclusion can be implemented, but are outside the scope of this article.

\begin{figure}[h]
\begin{tabular}{c}
$\Delta_{+}$
\\
\epsfxsize0.15\hsize\epsffile{figs/op2+0.ps} 
\\
$m = 0$ 
\end{tabular}
\hspace{0.75cm}
\begin{tabular}{c}
$\Delta_{-}$
\\
\epsfxsize0.15\hsize\epsffile{figs/op3+0.ps}
\\
$p = + 2$
\end{tabular}
\hspace{0.75cm} 
\begin{tabular}{c}
\\
\epsfxsize0.22\hsize\epsffile{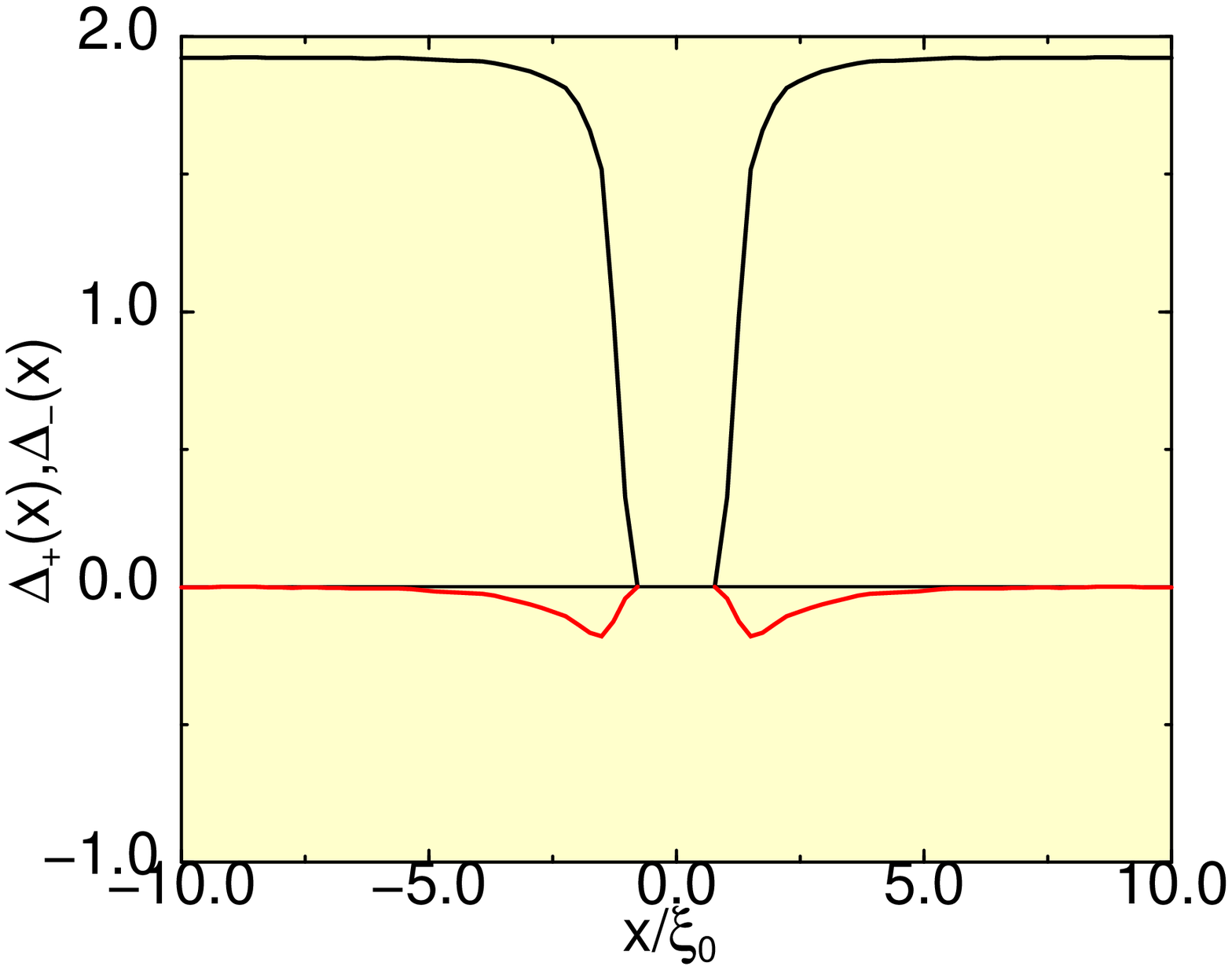} 
\\ 
\end{tabular}
\hspace{0.75cm} 
\begin{tabular}{c}
\\
\epsfxsize0.18\hsize\epsffile{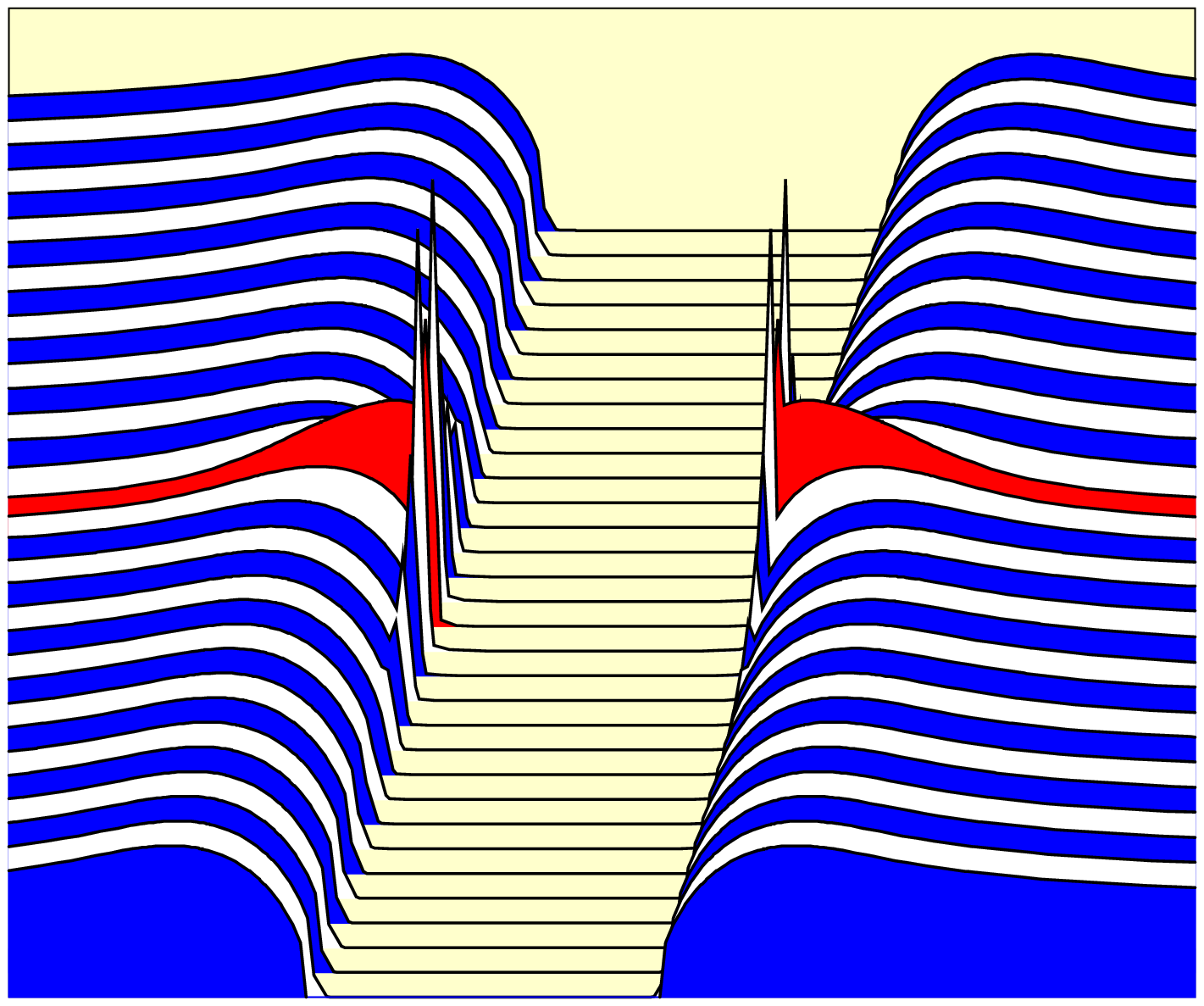} 
\\
\end{tabular}
\caption{
\label{Fig1_0}
The structure of a normal metallic inclusion embedded in the $(\hat{\vp}_x+i\hat{\vp}_y)$ phase. The inclusion is located in the center in column 1 and has a radius $r_{\text{incl}}=0.78\xi_0$.  The columns and notation are the same as that of Fig. \ref{Fig1_1}.  Note (i) the phase winding of the time-reversed order parameter (column 2)  and (ii) the shallow bound states induced by the inhomogeneous order parameter near the boundary of the metallic inclusion (column 4).
}
\end{figure}

\begin{figure}
\noindent
\centerline{
\begin{minipage}{0.28\hsize}
\epsfxsize0.7\hsize\epsffile{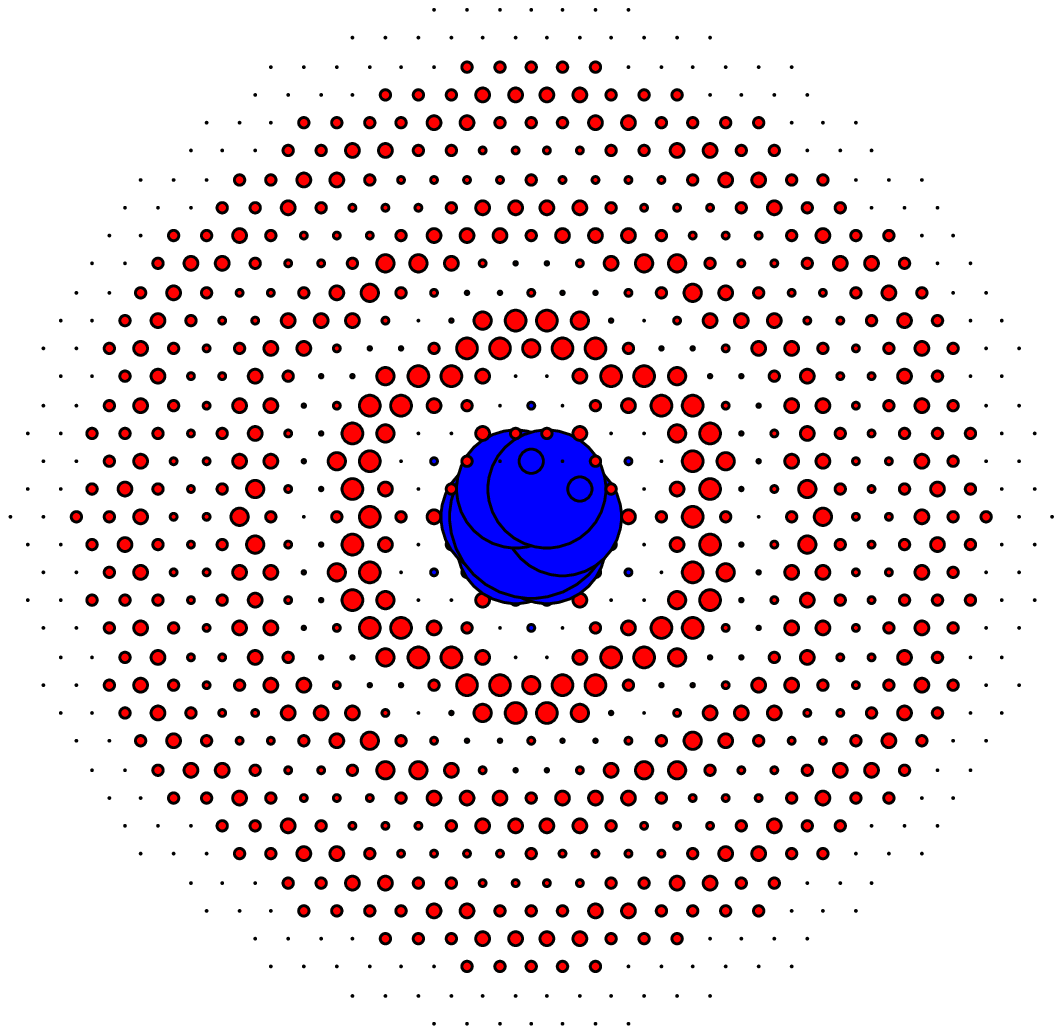}
\end{minipage}
\begin{minipage}{0.28\hsize}
\epsfxsize0.7\hsize\epsffile{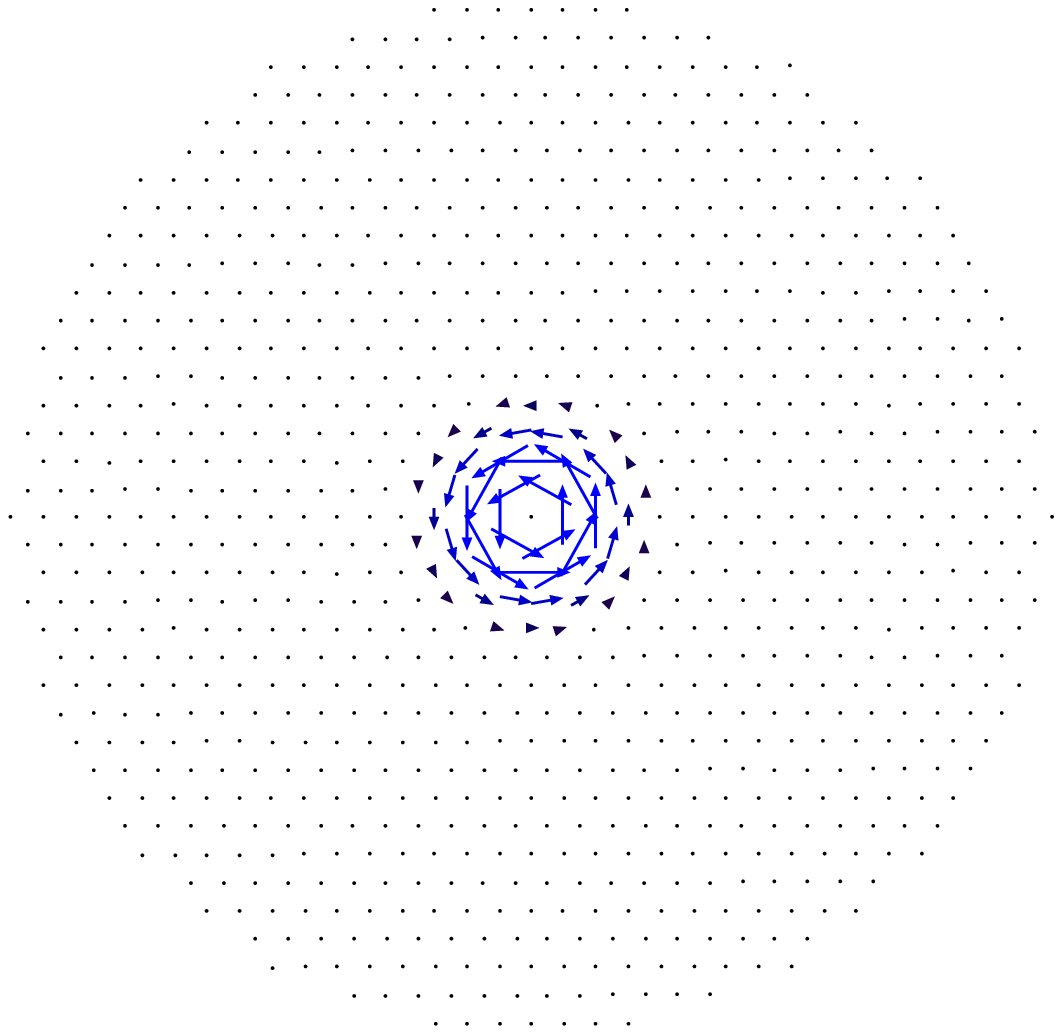}
\end{minipage} }
\centerline{
\begin{minipage}{0.28\hsize}
\epsfxsize0.7\hsize\epsffile{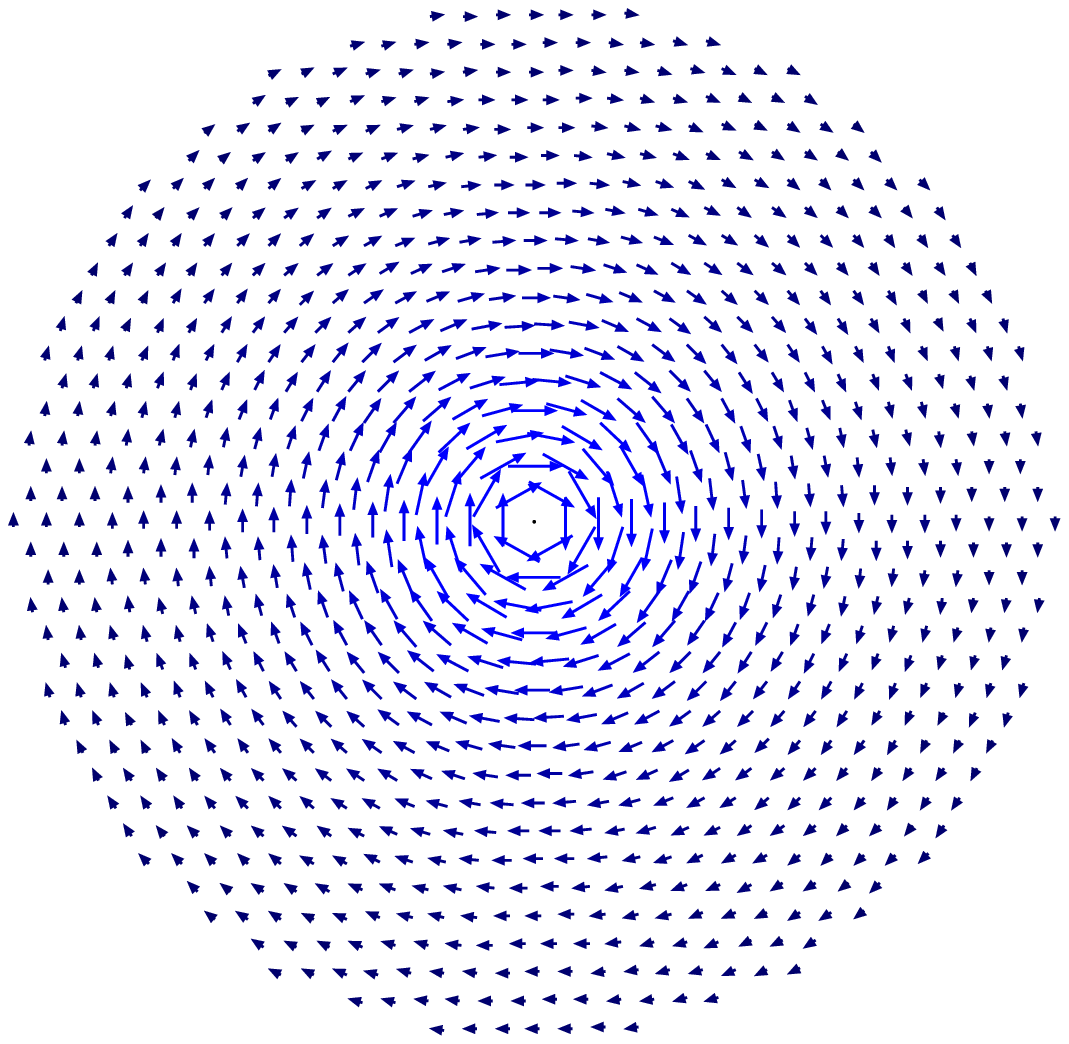}
\end{minipage}
\begin{minipage}{0.28\hsize}
\epsfxsize0.7\hsize\epsffile{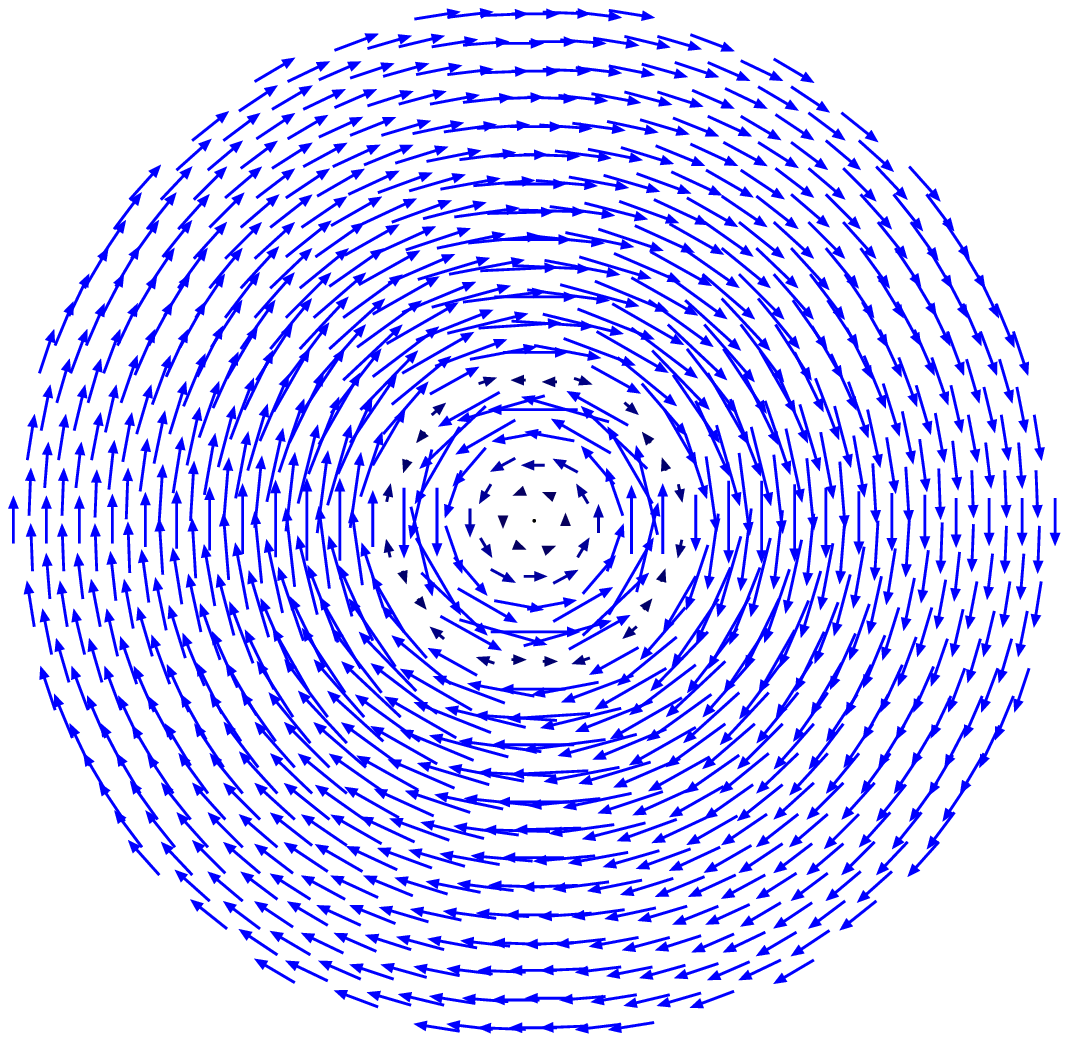}
\end{minipage}}
\caption{
\label{meissner}
Spontaneous local magnetic field (top-left) and current density (top-right) around a mesoscopic metallic inclusion with phase windings of $m=0,p=-2$. The change in sign of the induced field is indicated by change in color from blue to red. For comparison we show the current densities for a singly quantized vortex with $m=-1,p=1$ (bottom-left), and the doubly quantized vortex with $m=-2,p=0$ (bottom-right).
}
\end{figure}

Shown in column 2 of Fig. \ref{Fig1_0} is the time-reversed component of the order parameter, with a phase winding of $4\pi$, that nucleates near the metallic inclusion. Broken time-reversal symmetry is revealed by the appearance of spontaneous supercurrents that flow in and around the metallic inclusion as shown in the top-right panel of Fig. \ref{meissner}.
These currents generate a local magnetic field, $\vb = b(x,y) \hat\vz$, which oscillates in sign as one moves radially away from the defect on the scale of the coherence length, $\xi_0$, shown in top-left panel of Fig. \ref{meissner}. The spatial average of the field is zero, but the locally varying field is non-zero and should be observable, e.g. as broadening of the $\mu$SR linewidth below $T_c$. 
We also note that the current and field induced by a non-magnetic point impurity in a chiral p-wave superconductor \cite{cho89a} is similar to that of a mesoscopic metallic inclusion, however the origin and interpretation of the structure appears different.
Complex current and field distributions may also be generated in unconventional, but non-chiral, superconductors by magnetic and spin-orbit scattering impurities \cite{gra00}.

Finally we note that the leading higher order harmonic corrections resulting from discrete $C_4$ lattice symmetry are expected to be weak for a small cylindrically symmetric defect. The most important higher harmonic will be a  subdominant component with $p=-2$, but it is weaker than the leading p-wave harmonic with $p=+2$. All other harmonics can be neglected for small inclusions because the higher winding numbers will force the induced amplitude to be further from the vortex core and thus more effectively suppressed by the dominant $m=0$ order parameter. A possible exception may occur for large inclusions, $r_{\text{incl}}\gg\xi_0$. In this limit the harmonic with $p=6$ may be more important than the $p= - 2$ component.

\section{Conclusions}

In conclusion we have analyzed and carried out self-consistent calculations of the equilibrium structure and excitation spectrum of vortices in layered, spin-triplet, $p$-wave superconductors and superfluid \He\ films with spontaneously broken chiral symmetry in the ground state. 
We show that the cores of vortices contain the time-reversed phase with a local phase winding that differs by $\pm 2\times 2\pi$ depending on the chirality, i.e. $L^{\text{orb}}_z = \pm\hbar$, of the ground state.
This implies that vortices with equal but opposite phase winding are inequivalent, and thus the lower critical field for nucleation of vortices in layered, chiral p-wave superconductors depends on the relative orientation of the field and the orbital angular momentum of the ground-state pairs.
Axially symmetric doubly quantized vortices with zero phase winding in the core are predicted. A lattice of these vortices is energetically favored at sufficiently high fields for $-\vH  || \vL_z =\hbar \hat\vz$.
For the opposite phase winding, i.e. $\vH  || \vL_z = +\hat\vz$, the doubly quantized vortex contains four circulation quanta for the core amplitude. This vortex spontaneously breaks axial symmetry by dissociation of the core vorticity into four singly quantized vortices in the time-reversed order parameter induced in the core. A lattice of these vortices is also expected to be energetically favored at sufficiently high fields with a square lattice structure.
In addition to vortex states, inhomogeneities in the distribution of non-magnetic impurities, or metallic inclusions embedded in the ground state lead to spontaneous supercurrents and magnetic fields induced in the Meissner phase, which are localized near the inhomogeneity. 
These currents are carried by bound electronic states associated with an induced order parameter with phase winding of $4\pi$. The magnetic field and current distribution vary on the scale of the coherence length and are expected to be observable by sufficiently small local magnetic probes, or as a broadening of the $\mu$SR linewidth below $T_c$.

\ack
We acknowledge support from the National Science Foundation DMR-0805277 (JAS), the Center for Functional Nanostructures of the DFG (ME) and helpful discussions with Mikael Fogelstr\"om and Tomas L\"ofwander.  JAS acknowledges the hospitality of Kavli Institute for Theoretical Physics where this work was presented as part of the workshop on "Chiral Superconductivity", and ME acknowledges the hospitality of the Aspen Center for Physics.

\end{document}